\documentclass[preprint,authoryear,12pt]{elsarticle}

\usepackage{epsfig}

\usepackage{amssymb}
\usepackage{longtable}
\usepackage{natbib,afterpage}
\usepackage{graphicx,color}
\usepackage{amssymb,times,graphics, subfigure}
\usepackage[english]{babel}
\usepackage{txfonts}
\usepackage{subfigure}
\usepackage[rightcaption]{sidecap}
\usepackage{clearpage}

\def\Msun{\hbox{M$_{\odot}$}}               
\def\Rsun{\hbox{R$_{\odot}$}}               
\def\Lsun{\hbox{L$_{\odot}$}}               
\def\Rstar{\hbox{R$_{\star}$}}              
\def\Mdot{\hbox{$\dot{M}$}}               
\def\Teff{\hbox{$\rm{T}_{\rm eff}$}}            
\def\arcsec{\hbox{$^{\prime\prime}$}}
\def\arcmin{\hbox{$^{\prime}$}}

\def\deg{\hbox{$^\circ$}}       

\def\la{\mathrel{\mathchoice {\vcenter{\offinterlineskip\halign{\hfil$\displaystyle##$\hfil\cr<\cr\sim\cr}}}
{\vcenter{\offinterlineskip\halign{\hfil$\textstyle##$\hfil\cr<\cr\sim\cr}}}
{\vcenter{\offinterlineskip\halign{\hfil$\scriptstyle##$\hfil\cr
<\cr\sim\cr}}}
{\vcenter{\offinterlineskip\halign{\hfil$\scriptscriptstyle##$\hfil\cr><cr\sim\cr}}}}}

\bibpunct{(}{)}{;}{a}{}{,}

\journal{Advances in Space Research}

\newcount\longrefs
\def\aap{\ifnum\longrefs=1 {Astron.\ Astrophys.}\else 
                           {A\hbox{\rm \&}A}\fi}
\def\aapl{\ifnum\longrefs=1 {Astron.\ Astrophys.\ Lett.}\else 
                           {A\hbox{\rm \&}A}\fi}
\def\aapr{\ifnum\longrefs=1 {Astron.\ Astrophys.\ Rev.}\else 
                            {A\hbox{\rm \&}AR}\fi}
\def\aaps{\ifnum\longrefs=1 {Astron.\ Astrophys.\ Suppl.}\else 
                            {A\hbox{\rm \&}AS}\fi}
\def\aj{\ifnum\longrefs=1 {Astron.\ J.}\else 
                          {AJ}\fi} 
\def\ao{\ifnum\longrefs=1 {Applied Optics}\else 
                           {Appl.\ Opt.}\fi} 
\def\aspcs{\ifnum\longrefs=1 {Astron.\ Soc.\ Pacific Conf. Series}\else 
                           {ASP Conf.\ Ser.}\fi} 
\def\apj{\ifnum\longrefs=1 {Astrophys.\ J.}\else 
                           {ApJ}\fi} 
\def\apjl{\ifnum\longrefs=1 {Astrophys.\ J.\ Lett.}\else 
                            {ApJ}\fi} 
\def\aplett{\ifnum\longrefs=1 {Astrophys.\ J.\ Lett.}\else 
                            {ApJ}\fi} 
\def\apjs{\ifnum\longrefs=1 {Astrophys.\ J.\ Suppl.}\else 
                            {ApJS}\fi}
\def\apss{\ifnum\longrefs=1 {Astrophys.\ and Space Science}\else 
                            {Ap\hbox{\rm \&}SS}\fi}
\def\araa{\ifnum\longrefs=1 {Ann.\ Rev.\ Astron.\ Astrophys.}\else 
                            {ARA\hbox{\rm \&}A}\fi}
\def\azh{\ifnum\longrefs=1 {Astronomicheskii Zhurnal}\else 
                            {Astron.\ Zhur.}\fi}
\def\baas{\ifnum\longrefs=1 {Bull.\ Am.\ Astron.\ Soc.}\else 
                            {BAAS}\fi}
\def\bain{\ifnum\longrefs=1 {Bull.\ Astronom.\ Institutes Netherlands}\else
                            {Bull.\ Astr.\ Inst.\ Neth.}\fi}
\def\gca{\ifnum\longrefs=1 {Geochim.\ Cosmochim.\ Acta}\else 
                           {Geochim.\ Cosmochim.\ Acta}\fi}
\def\grl{\ifnum\longrefs=1 {Geophys.\ Res.\ Lett.}\else 
                           {Geoph.\ Res.\ Lett.}\fi}
\def\iaucirc{\ifnum\longrefs=1 {IAU Circulars}\else 
                          {IAU Circ.}\fi}
\def\ip{\ifnum\longrefs=1 {in press}\else 
                          {in press}\fi}
\def\jchemp{\ifnum\longrefs=1 {J.\ Chem.\ Phys.}\else 
                           {J.\ Chem.\ Phys.}\fi}  
\def\jcp{\ifnum\longrefs=1 {J.\ Chem.\ Phys.}\else 
                           {J.\ Chem.\ Phys.}\fi}  
\def\jgr{\ifnum\longrefs=1 {J.\ Geophys.\ Res.}\else 
                           {J.\ Geophys.\ Res.}\fi}  
\def\jmolspec{\ifnum\longrefs=1 {J.\ Mol.\ Spectrosc.}\else 
                           {J.\ Mol.\ Spectrosc.}\fi}  
\def\jqsrt{\ifnum\longrefs=1 {J.\ Quant.\ Spectrosc.\ Radiat.\ Transfer}\else 
                           {J.\ Quant.\ Spectrosc.\ Radiat.\ Transfer}\fi}  
\def\jrasc{\ifnum\longrefs=1 {J.\ Royal Astron.\ Soc.\ Canada}\else 
                           {JRAS Can.}\fi}  
\def\mnras{\ifnum\longrefs=1 {Mon.\ Not.\ Roy.\ Astron.\ Soc.}\else 
                             {MNRAS}\fi} 
\def\nat{\ifnum\longrefs=1 {Nature}\else 
                           {Nat}\fi}
\def\pasj{\ifnum\longrefs=1 {Pub.\ Astron.\ Soc.\ Japan}\else 
                            {PASJ}\fi} 
\def\pasp{\ifnum\longrefs=1 {Pub.\ Astron.\ Soc.\ Pacific}\else 
                            {PASP}\fi} 
\def\physscr{\ifnum\longrefs=1 {Physica Scripta}\else 
                            {Phys.\ Scrip.}\fi} 
\def\planss{\ifnum\longrefs=1 {Planetary \& Space Science}\else 
                            {Plan. \& Space Sci.}\fi} 
\def\procspie{\ifnum\longrefs=1 {Proc.\ SPIE}\else 
                            {Proc.\ SPIE}\fi} 
\def\qjras{\ifnum\longrefs=1 {Quarterly J.\ Royal Astron.\ Soc.}\else 
                            {QJRAS}\fi} 
\def\sa{\ifnum\longrefs=1 {Soviet Astron..}\else 
                               {Sov.\ Astron.}\fi}
\def\skytel{\ifnum\longrefs=1 {Sky \& Telescope}\else 
                            {Sky \& Tel.}\fi} 
\def\solphys{\ifnum\longrefs=1 {Solar Phys.}\else 
                               {Solar Phys.}\fi}
\def\ssr{\ifnum\longrefs=1 {Space Science Rev.}\else 
                               {Space\ Sci.\ Rev.}\fi}

\def\bibfiles{paper_LD}   
\def\fullreferences{\longrefs=1  \bibliographystyle{/lhome/leen/latex/bibtex/namebib}
             \bibliography{/lhome/leen/latex/bibtex/journals,\bibfiles}}

\begin{document}
\selectlanguage{english}

\begin{frontmatter}



\title{Late Stages of Stellar Evolution - Herschel's contributions\tnoteref{footnote1}}
\tnotetext[footnote1]{Herschel is an ESA space observatory with science instruments provided by European-led Principal Investigator consortia and with important participation from NASA.}


\author{Leen Decin\corref{cor}}
\address{Instituut voor Sterrenkunde,
             Katholieke Universiteit Leuven, Celestijnenlaan 200D, 3001 Leuven, Belgium. \\
Sterrenkundig Instituut Anton Pannekoek, University of Amsterdam, Science Park 904, NL-1098 Amsterdam, The Netherlands
}
\cortext[cor]{Corresponding author}
\ead{Leen.Decin@ster.kuleuven.be}




\begin{abstract}
Cool objects glow in the infrared. The gas and solid-state species that escape the stellar gravitational attraction of evolved late-type stars in the form of a stellar wind are cool, with temperatures typically $\la$1500\,K, and can be ideally studied in the infrared. These stellar winds create huge extended circumstellar envelopes with extents approaching $10^{19}$\,cm. In these envelopes, a complex kinematical, thermodynamical and chemical interplay determines the global and local structural parameters. Unraveling the wind acceleration mechanisms and deriving the complicated  structure of the envelopes is important to understand the late stages of evolution of $\sim$97\% of stars in galaxies as our own Milky Way. That way, we can also assess the significant chemical enrichment of the interstellar medium by the mass loss of these evolved stars. The \textit{Herschel Space Observatory} is uniquely placed to study evolved stars thanks to the excellent capabilities of the three infrared and sub-millimeter 
instruments on board: PACS, SPIRE and HIFI. In this review, I give an overview of a few important results obtained during the first two years of \textit{Herschel} observations in the field of evolved low and intermediate mass stars, and I will show how the \textit{Herschel} observations can solve some historical questions on these late stages of stellar evolution, but also add some new ones.

\end{abstract}


\end{frontmatter}

\parindent=0.5 cm


\section{Introduction} \label{SECT:Introduction}

The launch of the \textit{Herschel Space Observatory} \citep{Pilbratt2010A&A...518L...1P} in May 2009 can be marked as a milestone in the area of infrared astronomy. With a mirror of 3.5\,m in diameter, \textit{Herschel} studies the Universe in the far-infrared and sub-millimeter wavelength ranges (60--670\,$\mu$m) thanks to the capabilities of the three instruments on board: HIFI \citep{deGraauw2010A&A...518L...6D}, PACS \citep{Poglitsch2010A&A...518L...2P}, and SPIRE \citep{Griffin2010A&A...518L...3G}. 
The infrared (IR) and sub-millimeter regions of the spectrum are of great scientific interest, not only because it is here that cool objects (10--1000\,K) radiate the bulk of their energy, but also because of its rich variety of diagnostic atomic, ionic, molecular and solid-state spectral features. Measurements at these wavelengths permit determination and evaluation of the physical processes taking place in astronomical sources, establishing the energy balance, temperature, abundances, density and velocity.

The \textit{Herschel} instruments enable the astronomers to study the coldest and dustiest objects in space, among which star forming regions, young stars, evolved stars, dusty galaxies, \ldots In this review, we will focus on the late stages of evolution of low and intermediate-mass stars (LIMS, M$\sim$0.6--8\,\Msun), the mass interval to which most of the stars in the Universe belong. Using a standard Salpeter initial mass function (IMF), 97\% of all stars born will go through the so-called Asymptotic Giant Branch (AGB) phase, where mass loss dominates the stellar evolution. More massive stars will lose a significant fraction of their mass in the so-called supergiant phase. The mass loss of AGB stars dominates the total mass of the interstellar medium (ISM) and almost half of the heavy elements returned to the ISM originates in the stellar interiors of these stars. 

The detailed physical description of the mass-loss process for late-type stars is, however, still unknown after several decades of studies. The generally accepted idea on the mass-loss mechanism for asymptotic giant branch (AGB) stars is based on pulsations and radiation pressure on newly formed dust grains. However, oxygen-rich AGB stars suffer from the so-called 'acceleration deficit' dilemma, which   states that mass-loss rates due to the formation of silicate dust alone are orders of magnitude smaller than  observed ones \citep{Woitke2006A&A...460L...9W}. The formation of large grains in the inner wind region might solve this riddle \citep{Hoefner2008A&A...491L...1H, Norris2012Natur.484..220N}. The proposed models for the mass loss of AGB stars are very unlikely to be applicable to red supergiants (RSGs) since they are irregular variables with small amplitudes. Processes linked to convection, chromospheric activity or rotation might play an important role as trigger for the mass loss in RSGs. \citet{
Josselin2007A&A...469..671J} proposed that turbulent pressure generated by convective motions, combined with radiative pressure on molecular lines, might initiate mass loss, but also Alfv\'en waves generated by a magnetic field might contribute \citep{Hartmann1984ApJ...284..238H}.

\begin{figure*}[htp]
 \includegraphics[width=\textwidth]{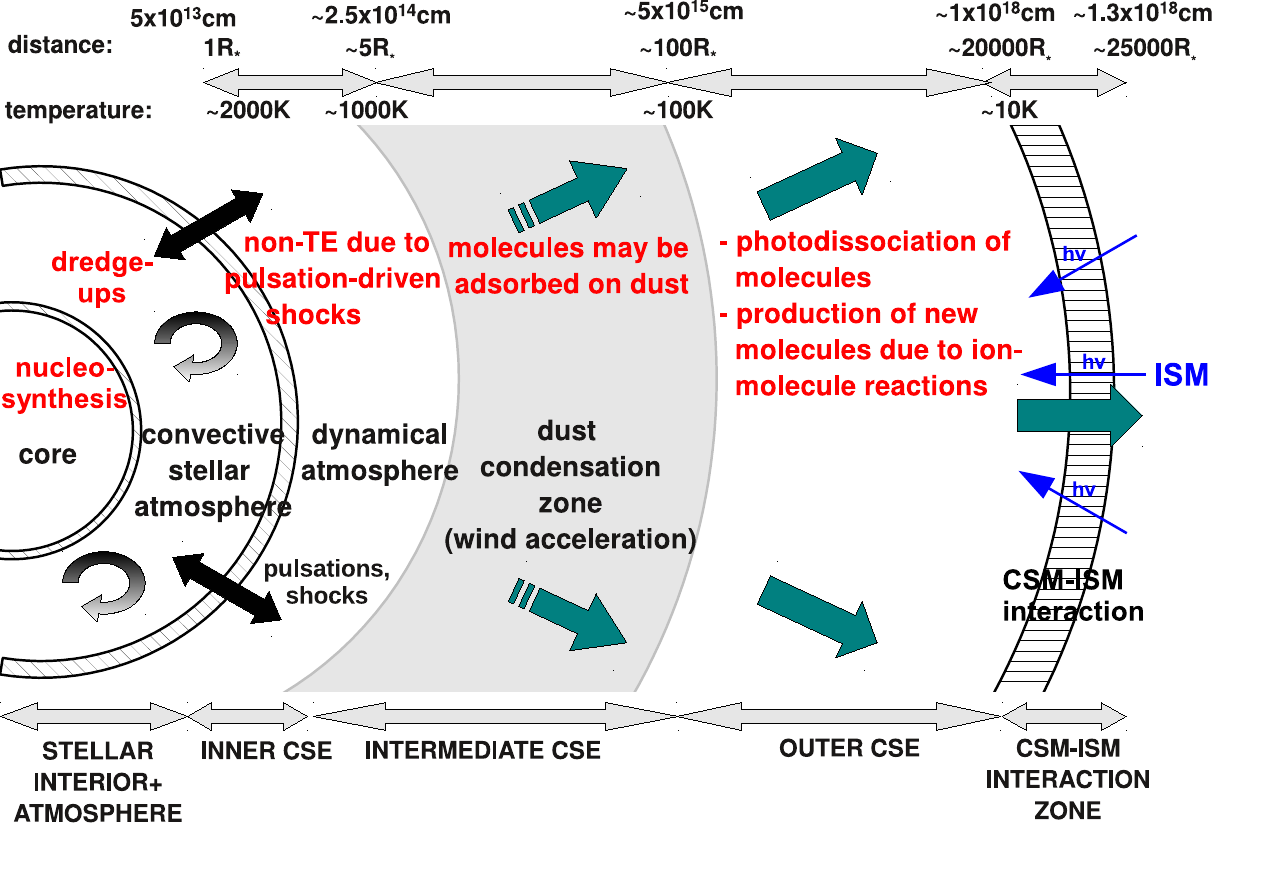}
\caption{Schematic drawing (not to scale) of an AGB star. Pulsations and dust formation are the two key ingredients for the generation of a stellar wind, with typical terminal velocities between 10--30\,km/s and  mass-loss rates between $10^{-8}$ and $10^{-4}$\,\Msun/yr.
Several chemical processes are indicated (in red) at the typical temperature and radial distance from the star  where they occur ['TE' stands for thermal equilibrium]. The penetration of interstellar ultraviolet photons in the outer envelope is shown in blue. Understanding the chemical processes in AGBs is crucial to know the chemical content of the (unprocessed) ISM. The results described in Sect.~\ref{SEC:GEOMETRY} and Sect.~\ref{SEC:CHEMISTRY} will cover subsequently the different geometrical regions displayed in this figure.}
\label{FIG:scheme_AGB}
\end{figure*}

Several guaranteed time and open time \textit{Herschel} proposals (will) focus on different aspects of these late stages of stellar evolution dominated by mass loss. E.g.\ large maps made with the PACS and SPIRE photometers will probe the infrared signature of dust emitted by evolved stars and supernovae in the Magellanic Clouds  \citep{Meixner2010A&A...518L..71M}. The goal is to assess the life cycle of matter in (external) galaxies by deriving general properties of the circumstellar material (CSM) surrounding these evolved stars. However, the details on the onset of the mass loss, on the mass-loss history, and on the chemistry, kinematics and thermodynamics in the complex outflows (see Fig.~\ref{FIG:scheme_AGB}) can only be assessed for the most nearby evolved stars (closer than few kpc). Unravelling these physical and chemical details will enhance our knowledge on the mass loss process in evolved stars, and will hopefully enable us to  derive some general quantities/laws applicable to more distant stars.
  
In this review, I will give an overview of some of the new scientific insight gained thanks to \textit{Herschel} after being 2.5\,yr into mission. The focus of Sect.~\ref{SEC:GEOMETRY} is on new geometrical and dynamical insights in the evolved stars' circumstellar envelopes (CSEs) as created by their stellar winds. Most of the constraints will come from PACS and SPIRE images, with photometer bands at 70, 100, and 160\,$\mu$m, and 250, 350, and 500\,$\mu$m respectively.  Sect.~\ref{SEC:CHEMISTRY} illustrates some spectroscopic results obtained with the high-spectral resolution capabilities of HIFI (157--212\,$\mu$m and 240--625\,$\mu$m, $R$ as high as $10^7$) and the medium-spectral resolution PACS (60--210\,$\mu$m, $R=1000-5000$) and SPIRE (194--672\,$\mu$m, $R=40-1000$) spectrometers. In each of these two sections, the examples will cover subsequently the different geometrical regions displayed in Fig.~\ref{FIG:scheme_AGB}. In Sect.~\ref{SECT:QUESTIONS}, I will discuss the answers on some `old' 
astronomical questions concerning these late stages of evolution of LIMS, but will also add some new questions to which the answer is still unclear. Some future perspectives are given in Sect.~\ref{SECT:PERSPECTIVES}, and some conclusions are drawn in Sect.~\ref{SECT:Conclusions}.

\section{New geometrical and dynamical insights from \textit{Herschel}}\label{SEC:GEOMETRY}

Despite the importance of the mass-loss process in terminating the AGB phase and in replenishing the ISM with newly-produced elements, the nature of the mass loss is still not well understood; in particular, one cannot predict the mass-loss rate for a star of given properties.

While for modeling purposes, the envelope is often assumed to be spherically symmetric and formed by a constant mass-loss rate, observations show more and more evidence contradicting these two basic assumptions.
CO and scattered light observations of (post) AGB objects show that winds from late-type
stars are far from being smooth. The shell structures found around, e.g., AFGL 2688 \citep{Sahai1998ApJ...493..301S} and IRC\,+10216 \citep{Mauron1999A&A...349..203M} indicate that the outflow has quasi-periodic oscillations, with density contrast corresponding to a factor $\sim$100--1000 \citep{Schoier2005A&A...436..633S, Decin2006A&A...456..549D}. These mass-loss modulations on a time scale of a few hundred years may be due to a complex feedback between hydrodynamics, gas/dust drift, and gas-solid chemistry \citep{Simis2001A&A...371..205S}. At the tip of the AGB, the mass ejection process becomes very strong, with mass-loss rates as high as 10$^{-4}$\,\Msun/yr. 

\textit{Herschel} data are now adding more and more complexity to this geometrical picture. In this next subsections, recent \textit{Herschel} results will be described covering gradually the envelope from the inner to the outermost structures.

\subsection{Inner envelope structure} \label{SECT:GEO_INNNER}

The PACS and SPIRE image capabilities for nearby evolved stars are tremendous. Thanks to their high spatial resolution (of $\sim$1\arcsec), one is able to map the structure of the inner envelope surrounding these evolved stars in quite some detail. Deducing if the envelope shows deviations from a smooth 1D envelope structure, created by a constant mass-loss rate, is however not so straightforward due to the complex point-spread function (PSF) of the PACS photometer and the fact that the cool central target (with typical temperatures between 2000--3500\,K) can be quite bright at these infrared wavelengths.

\begin{figure*}[htp]
 \includegraphics[width=0.32\textwidth]{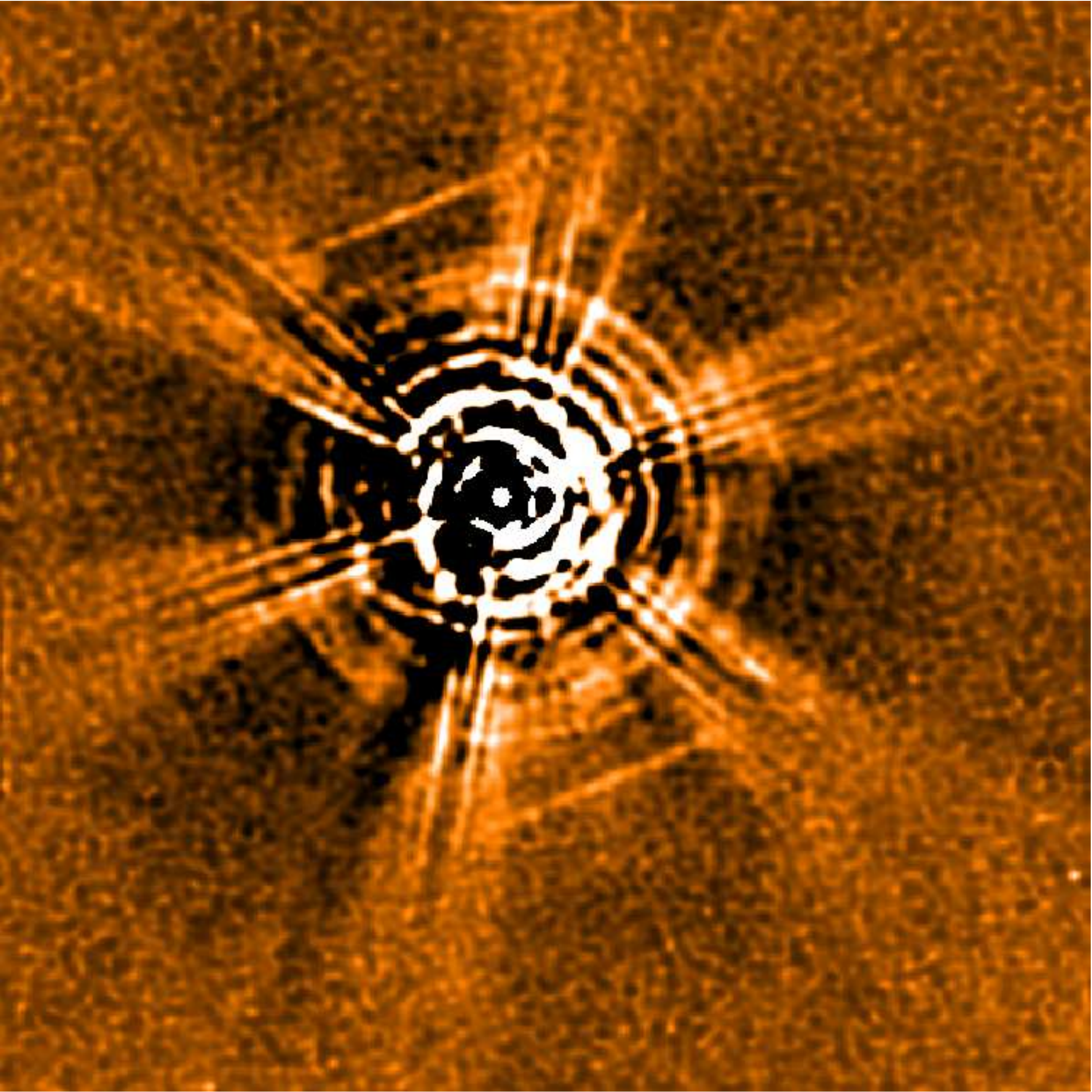}
 \includegraphics[width=0.32\textwidth]{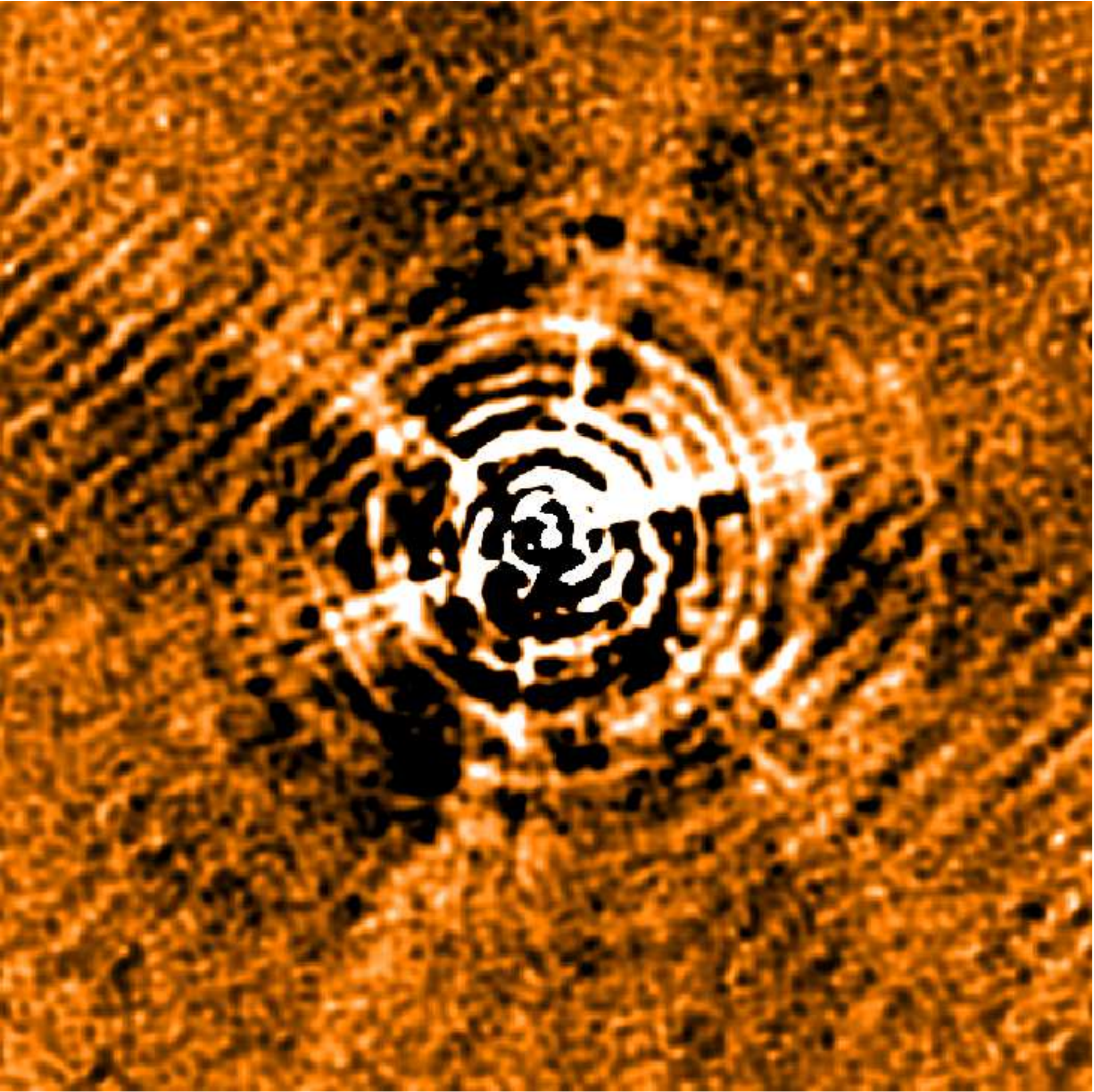}
 \includegraphics[width=0.32\textwidth]{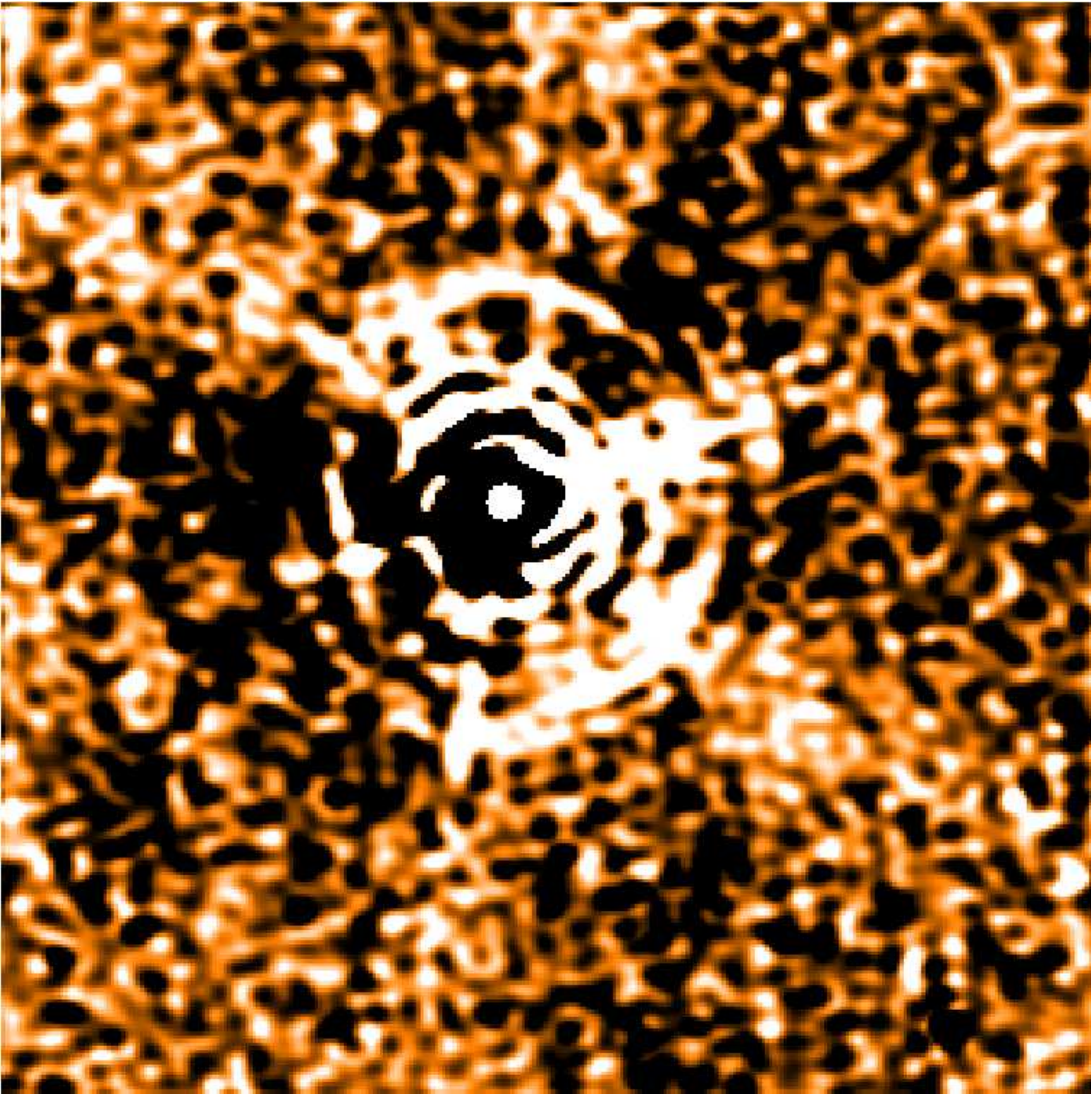}
\caption{PACS images of IRC\,+10216 at 70\,$\mu$m (left), 100\,$\mu$m (middle) and 160\,$\mu$m (right) after subtraction of the smooth halo of the circumstellar envelope \citep{Decin2011A&A...534A...1D}. The six radial spikes are due to the support spider structure of the secondary mirror. The field-of-view is 16\arcmin$\times$11\arcmin.} 
\label{FIG:CWLEO_PACS_70}
\end{figure*}

\begin{figure*}[!htp]
     \begin{minipage}[t]{.062\textwidth}
    \end{minipage}
     \begin{minipage}[t]{.561\textwidth}
        \centerline{\textbf{\phantom{AAAAAAAA}PACS 70\,$\mu$m}}
    \end{minipage}
    \hfill
    \begin{minipage}[t]{.379\textwidth}
        \centering{\textbf{PACS 70\,$\mu$m smooth envelope structure subtracted}}
    \end{minipage}

   \begin{minipage}[t]{.062\textwidth}
        \vspace*{-3cm}
        \centering{\textbf{\large{R Dor}}}
    \end{minipage}
   \begin{minipage}[t]{.469\textwidth}
        \centerline{\resizebox{\textwidth}{!}{\includegraphics{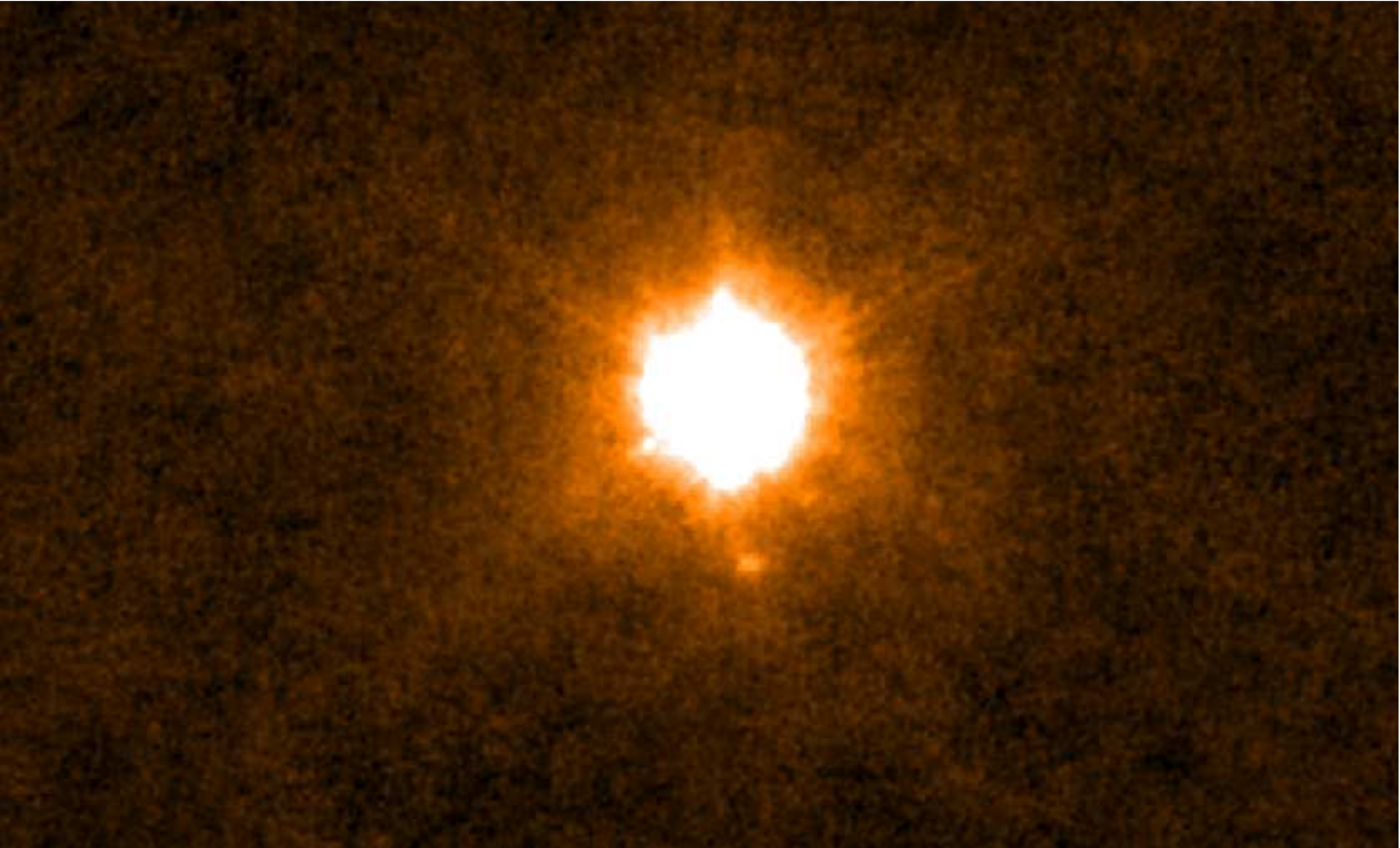}}}
    \end{minipage}
    \hfill
    \begin{minipage}[t]{.469\textwidth}
        \centerline{\resizebox{\textwidth}{!}{\includegraphics{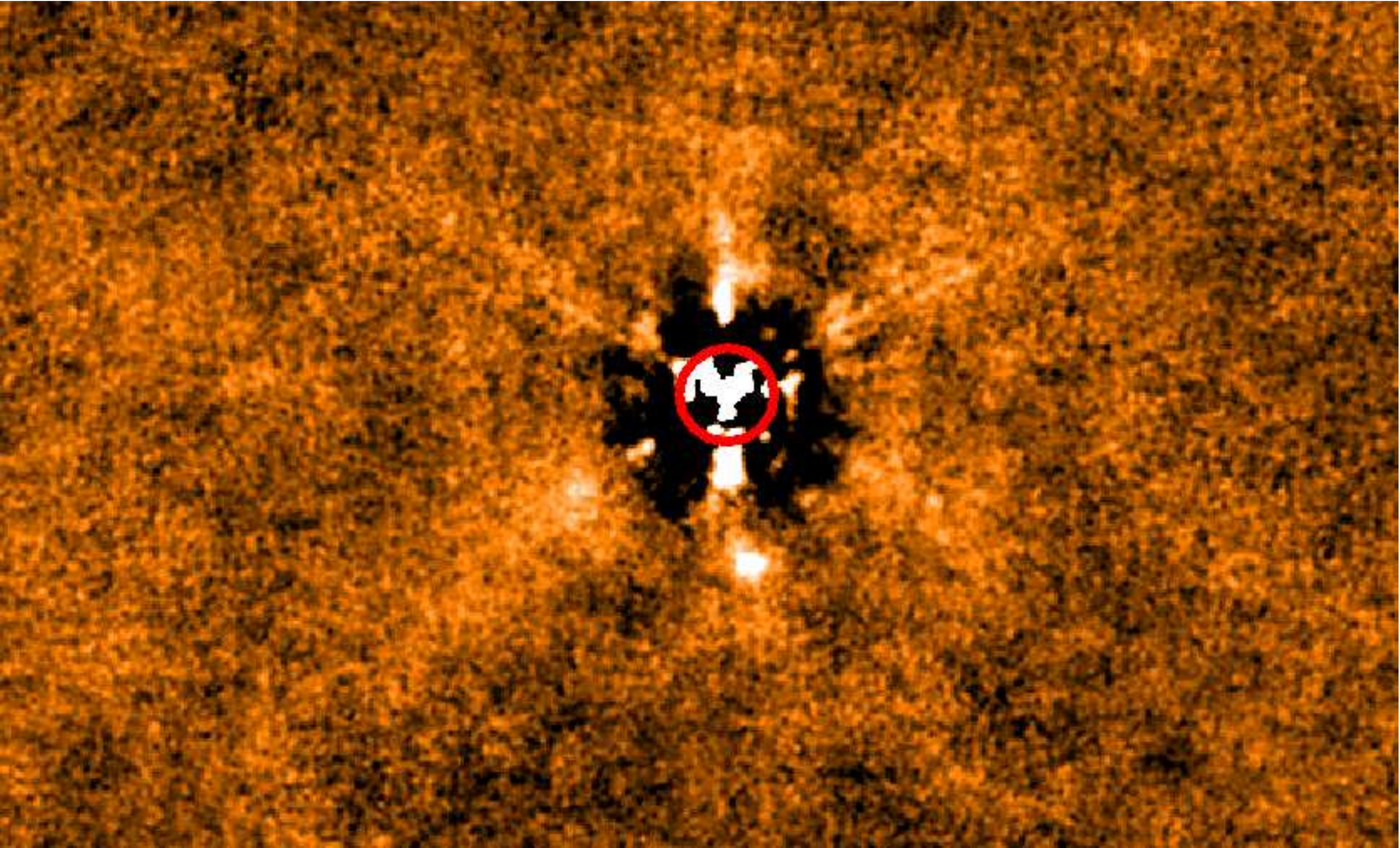}}}
    \end{minipage}
 
   \begin{minipage}[t]{.062\textwidth}
        \vspace*{-3cm}
        \centering{\textbf{\large{$\chi$ Cyg}}}
    \end{minipage}
   \begin{minipage}[t]{.469\textwidth}
        \centerline{\resizebox{\textwidth}{!}{\includegraphics{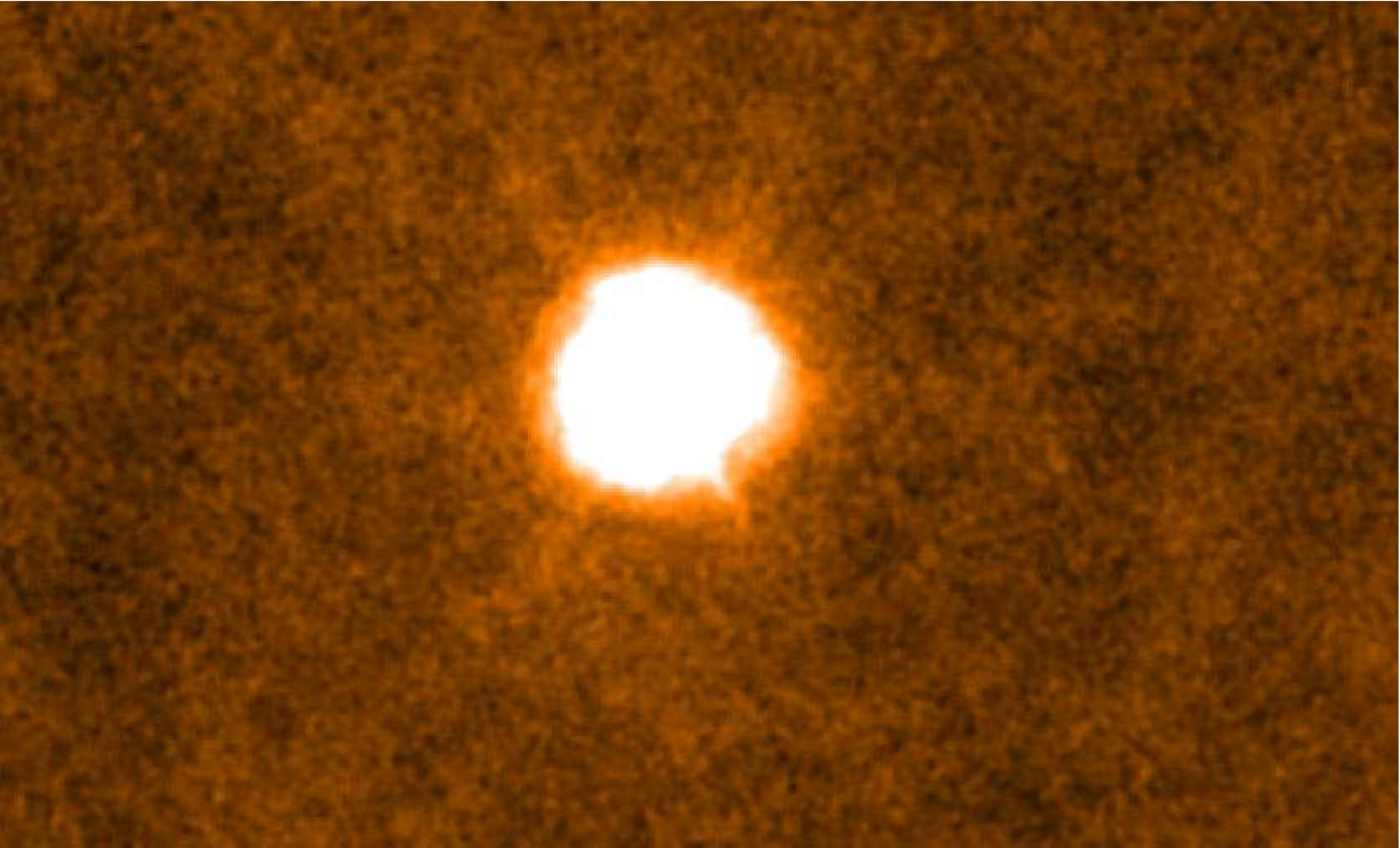}}}
    \end{minipage}
    \hfill
    \begin{minipage}[t]{.469\textwidth}
        \centerline{\resizebox{\textwidth}{!}{\includegraphics{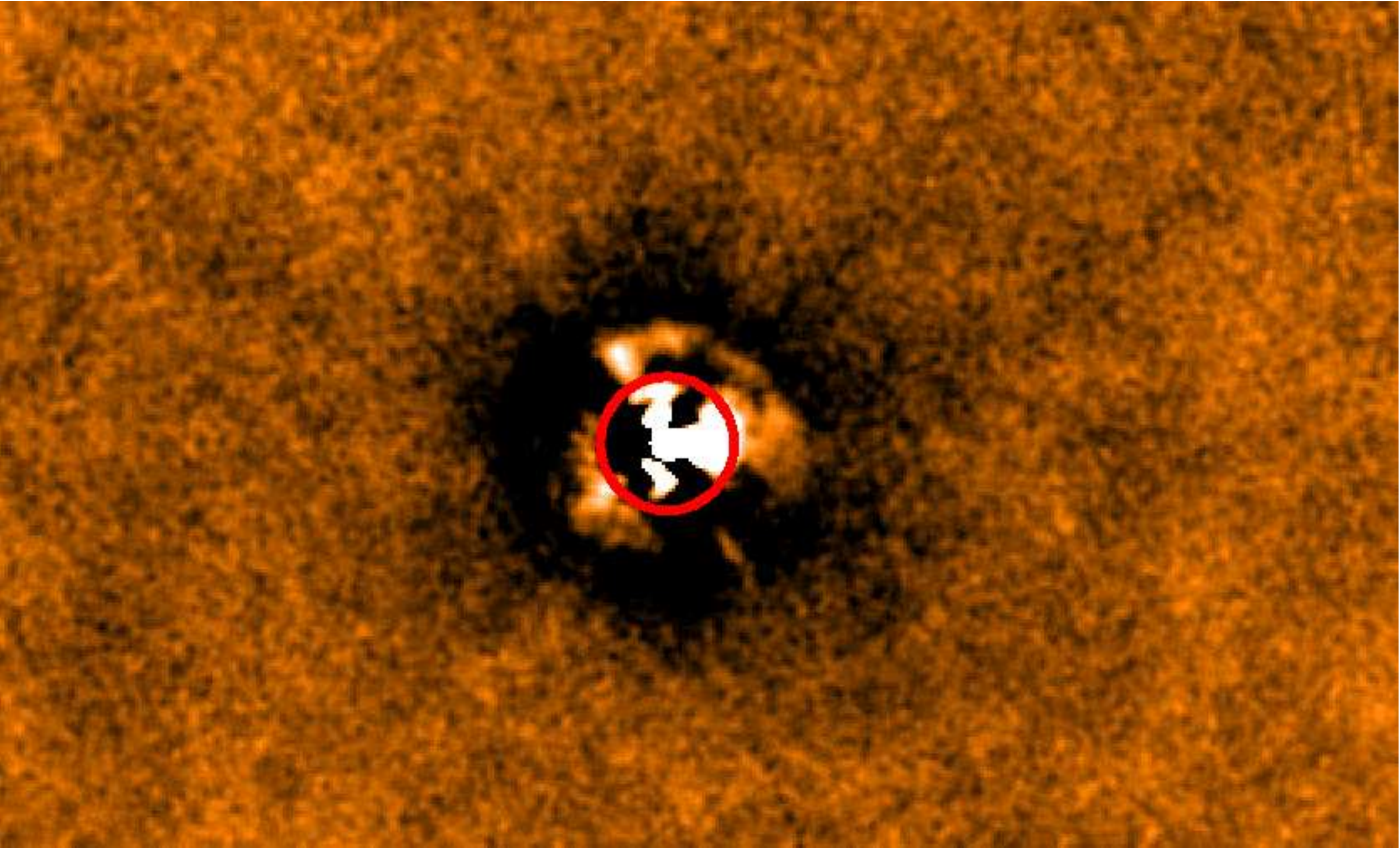}}}
    \end{minipage}
 
   \begin{minipage}[t]{.062\textwidth}
        \vspace*{-3cm}
        \centering{\textbf{\large{$\mu$ Cep}}}
    \end{minipage}
   \begin{minipage}[t]{.469\textwidth}
        \centerline{\resizebox{\textwidth}{!}{\includegraphics{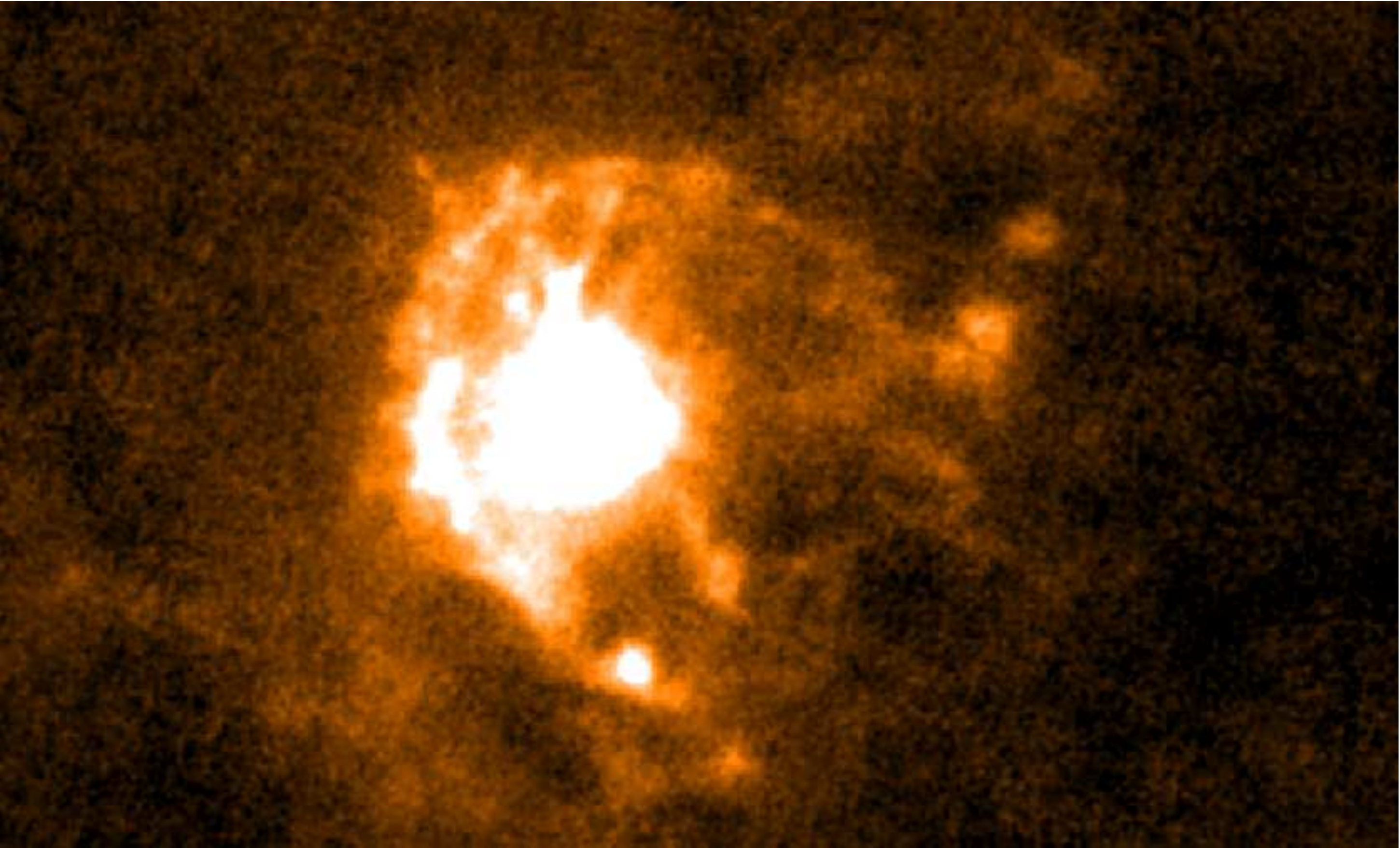}}}
    \end{minipage}
    \hfill
    \begin{minipage}[t]{.469\textwidth}
        \centerline{\resizebox{\textwidth}{!}{\includegraphics{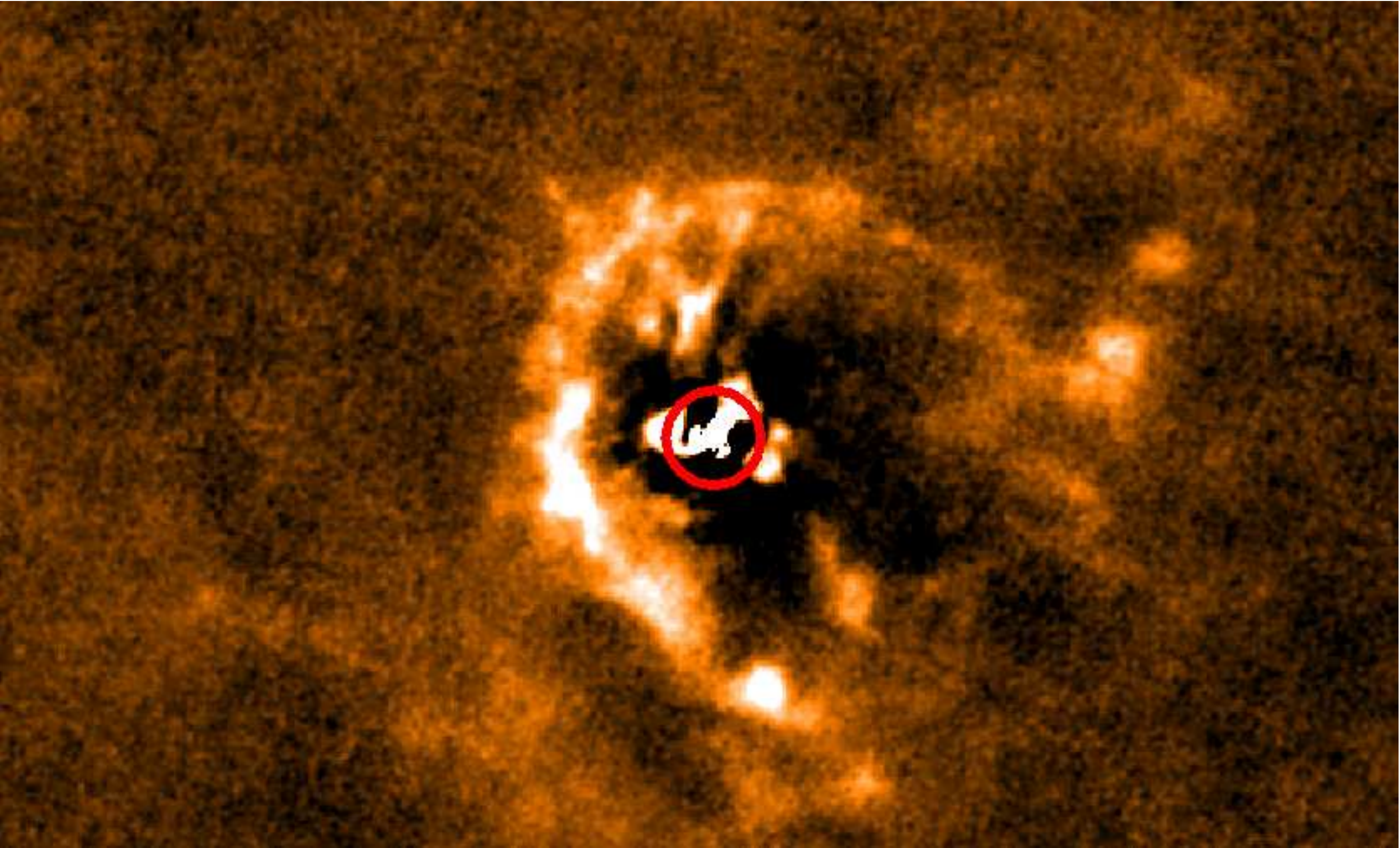}}}
    \end{minipage}
 
   \begin{minipage}[t]{.062\textwidth}
        \vspace*{-3cm}
        \centering{\textbf{\large{$\alpha$ Ori}}}
    \end{minipage}
   \begin{minipage}[t]{.469\textwidth}
        \centerline{\resizebox{\textwidth}{!}{\includegraphics{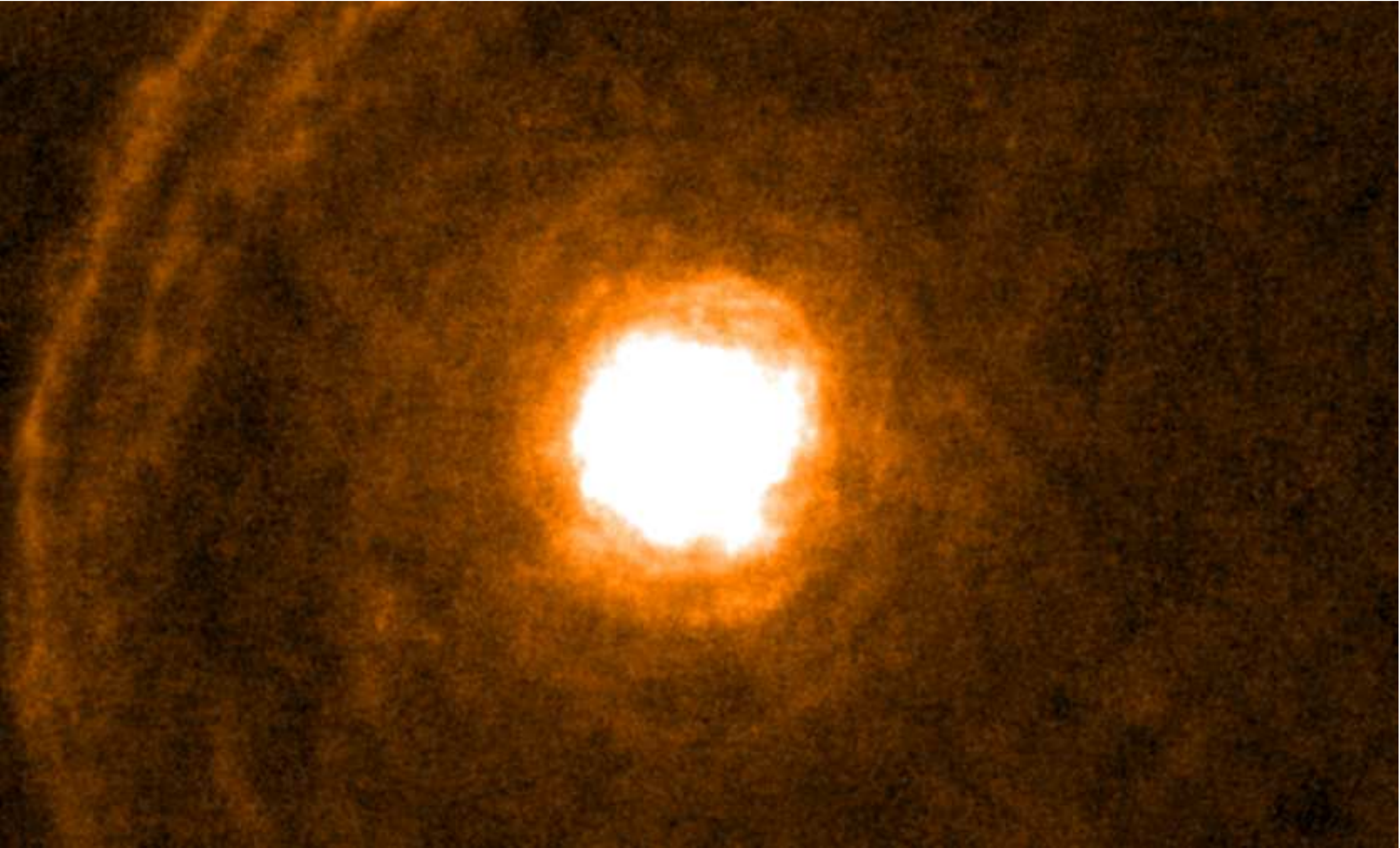}}}
    \end{minipage}
    \hfill
    \begin{minipage}[t]{.469\textwidth}
        \centerline{\resizebox{\textwidth}{!}{\includegraphics{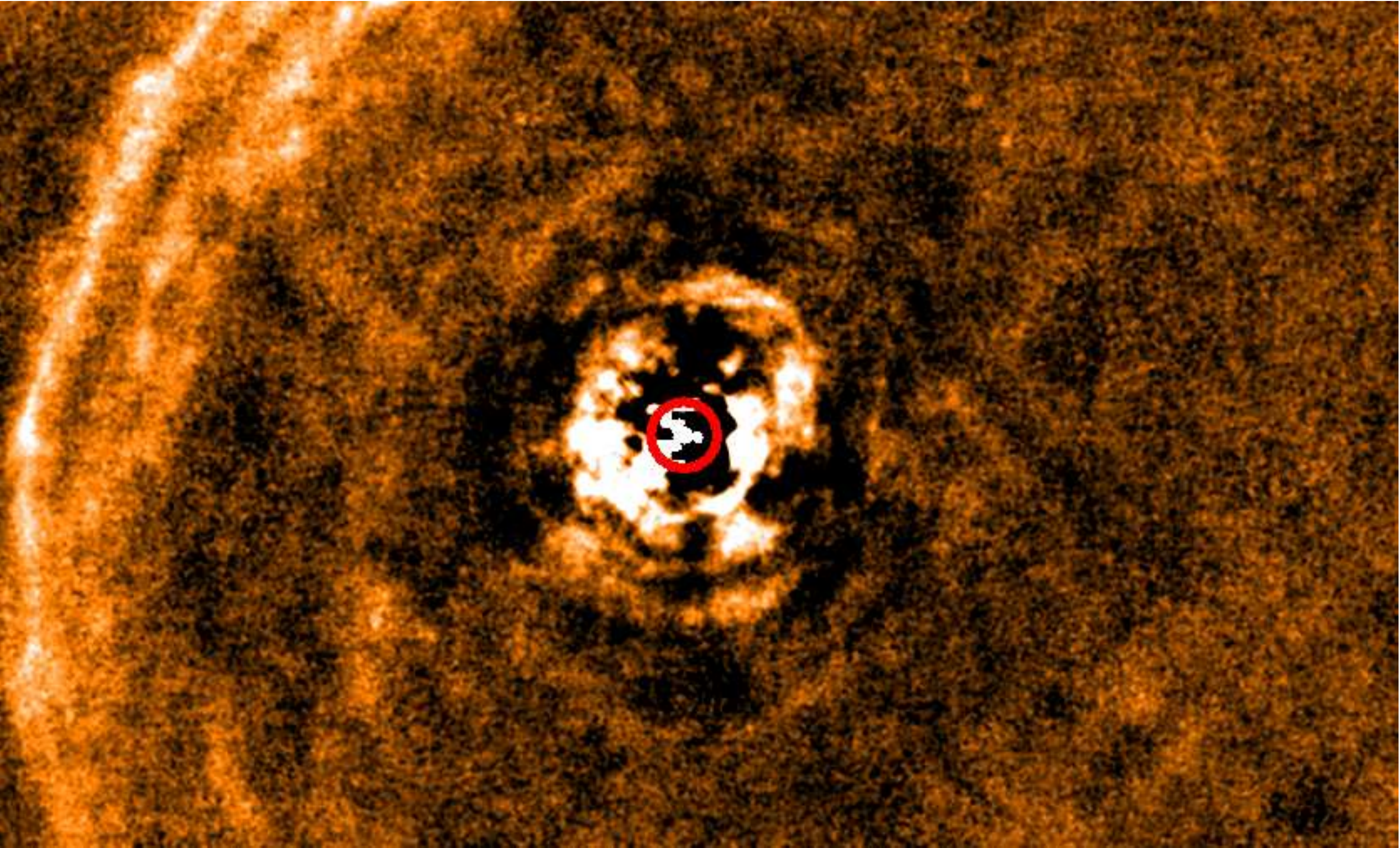}}}
    \end{minipage}
 \caption{PACS images of four evolved stars before (left) and after (right) subtraction of the smooth halo of the circumstellar envelope. North is to the top, east is to the left. The six radial spikes (visible in the image of R~Dor) are due to the support spider structure of the secondary mirror. The red circle with a radius of 15\arcsec\ marks the region where some PSF artifacts are still visible. These data were obtained in the framework of the \textit{Herschel} MESS Guaranteed Time Key Programme \citep{Groenewegen2011A&A...526A.162G}. A full analysis of the inner envelope structure of some tens of stars will be presented by Royer et al. (\textit{in prep.})}
 \label{FIG:inner_envelope}
\end{figure*}

To emphasize the shell morphology in the inner envelope, one can remove the extended envelope halo by subtracting a smooth, azimuthally averaged profile represented by a power law $r^{-\alpha}$  \citep[see Fig.~\ref{FIG:CWLEO_PACS_70},][]{Decin2011A&A...534A...1D}. The method has been tested in detail (Royer et al., \textit{in prep.}). An illustration for a point-like source is shown in the upper panels of Fig.~\ref{FIG:inner_envelope}: for the oxygen-rich semi-regular pulsating AGB star R~Dor, experiencing a low mass-loss rate of $\sim 1.2 \times 10^{-7}$\,\Msun/yr \citep{Schoier2004A&A...422..651S} no indications can be found for deviations from a smooth spherically symmetric envelope at scales $>$15\arcsec. 

The images of the three other targets shown in Fig.~\ref{FIG:inner_envelope} and the image of IRC\,+10216 (see Fig.~\ref{FIG:CWLEO_PACS_70}) testify to a much more complex envelope structure.
\paragraph{IRC\,+10216} is the nearest carbon-rich AGB star at a distance of $\sim$150\,pc \citep{Crosas1997ApJ...483..913C}. \citet{Mauron1999A&A...349..203M} have detected multiple, almost concentric, shells (or arcs) out to a distance of 55\arcsec\ in radius on top of the smooth extended envelope. The \textit{Herschel} images show the discovery of multiple dust shells until 320\arcsec. \citet{Decin2011A&A...534A...1D} argue that the origin of the shells is related to non-isotropic mass-loss events and clumpy dust formation.

\paragraph{$\chi$ Cyg} is the prototype of a S-type (C/O ratio slightly lower than 1) Mira variable with a period of 408 days at a distance of $\sim$150 pc and a mass-loss rate of $\sim 5 \times 10^{-7}$\,\Msun/yr \citep{Schoier2011A&A...530A..83S}. The PACS image shows some clear intensity enhancements in the north-western and south-eastern direction. \citet{Ragland2006ApJ...652..650R} reported the detection of asymmetric brightness distributions  using the IOTA interferometer in the $H$-band, with a angular resolution of 5--10\,mas, probably arising from just above the stellar photosphere. This is confirmed by \citet{Tatebe2006ApJ...652..666T}, who interpreted their interferometric data in terms of a steady excess of dust emitted to the east, or alternatively, as being due to a hot spot on the east side of the star that illuminates a symmetric dust shell in an asymmetric way. According to \citet{Tatebe2006ApJ...652..666T} a companion is unlikely to be the cause of the reported asymmetry. The \textit{
Herschel}/HIFI line  profiles of the H$_2$O, SiO and HCN show a consistent weak asymmetry, which are  attributed to small-scale deviations from spherical symmetry and the presence of a (weakly) clumped medium \citep{Schoier2011A&A...530A..83S}.

\paragraph{$\mu$ Cep} is an oxygen-rich supergiant at a distance of 390\,pc \citep{Cox2012A&A...537A..35C}. The PACS 70\,$\mu$m image is dominated by the bow shock structure situated at $\sim$50\arcsec\ (see Sect.~\ref{SECT:bow shock}). Some asymmetries at $\sim$20\arcsec\ away in the envelope are seen in the PACS image, although it is not clear if this is related to a turbulent mass-loss history or due to the large instabilities in the bow shock.

\paragraph{$\alpha$ Ori} is also an oxygen-rich supergiant at a distance of 197\,pc \citep{Harper2008AJ....135.1430H}. Many detailed studies have already described some characteristics of the complex atmosphere, chromosphere and dusty envelope of Betelgeuse. Irregular structures have been reported within 2.5\arcsec\ from the central target \citep[e.g.][]{Kervella2011A&A...531A.117K}. The \textit{Herschel} images show the first evidence for a high degree of clumpiness of the material lost by the star beyond 15\arcsec, which even persists until the material collides with the ISM \citep{Decin2012A}. Very pronounced asymmetries are visible within 110\arcsec\ from the central star, although some weaker flux enhancements are visible until $\sim$300\arcsec\ away.  The typical angular extent ranges from $\sim$10--90\deg.

Indeed, five examples of evolved AGB and supergiant stars, and  five different mass-loss histories. There is one common denominator: the emission we see is (almost completely) caused by dust -- there might be some minor contribution from some atomic or molecular lines in the photometer bands as [O\,I] at 63\,$\mu$m. This means that we are the direct witnesses of the role of dust in creating the envelope structure around these evolved stars. In the \textit{Herschel}/MESS GTKP some 78 AGB and supergiant stars are observed; other open and guaranteed time proposals will image some dozens of other nearby evolved stars. It is clear that these results will revolutionize our ideas on the wind creation in these targets, and that these images form the perfect ground for follow-up observations with ALMA.

\subsection{Wind acceleration}

\begin{figure}[htp]
 \includegraphics[angle=180,width=\textwidth]{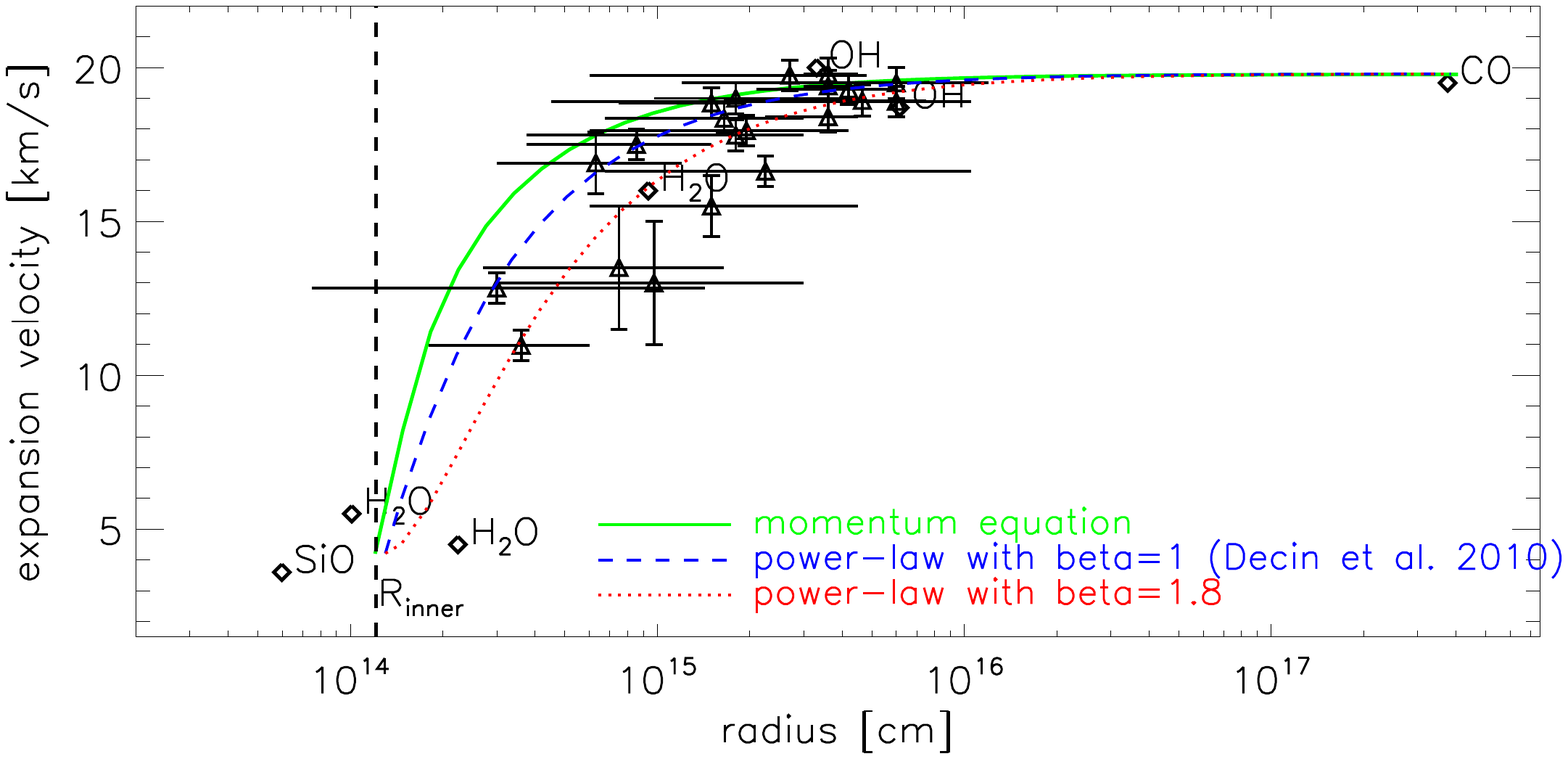}
\caption{Velocity profile of IK~Tau as published in \citet{Decin2010A&A...521L...4D}. Velocity data are obtained from mapping of maser emission: SiO \citep{Boboltz2005ApJ...625..978B}, H$_2$O \citep{Bains2003MNRAS.342....8B}, and OH \citep{Bowers1989ApJ...340..479B}. The CO expansion velocity derived from ground-based CO J=1--0 data is also indicated \citep{Decin2010A&A...516A..69D}. The triangles show the place in the envelope where the line formation is highest for the HIFI data presented in \citet{Decin2010A&A...521L...4D}, including $^{12}$CO, $^{13}$CO, H$_2^{16}$O, H$_2^{17}$O, H$_2^{18}$O, $^{28}$SiO, $^{29}$Si), $^{30}$SiO, HCN, and SO lines. The horizontal bars show the minimum and maximum radial distance for the line formation of each individual transition. The vertical bars show the uncertainty on the observed line widths.
The expansion velocity deduced from solving the momentum equation (assuming extinction efficiencies for spherical dust particles with a main component being amorphous olivine) is shown by the full green line. The dashed blue line represents a power law (Eq.~\ref{Eq:beta}) with $\beta$\,=\,1. For comparison, an even smoother expansion velocity structure with $\beta$\,=\,1.8 is shown with the red dotted line. The vertical dashed black line indicates the dust condensation radius R$_{\rm inner}$.  The velocity at R$_{\rm inner}$ is assumed to be equal to the local sound velocity.
}
\label{FIG:IKTAU_VELOCITY}
\end{figure}

The formation of dust species is thought to play a crucial role for the onset of the wind acceleration in AGB stars. In case of the supergiants, the role of the dust species for the wind acceleration is absolutely not clear. But even in case of  AGB stars, it is  not yet understood how the wind can be driven in an oxygen-rich environment, where the glassy character of the oxides and pure 
(magnesium-rich) silicates --- the species which might form close to the star --- prevents them from gaining enough momentum \citep{Woitke2006A&A...460L...9W}\footnote{When solving the momentum equation shown in Fig.~\ref{FIG:IKTAU_VELOCITY}, the silicate dust composition as derived by \citet{Justtanont1992ApJ...389..400J} was assumed. The main component is amorphous olivine 
(Mg$_x$Fe$_{1-x}$SiO$_4$).}. The formation of both carbon and silicate grains  \citep{Hoefner2007A&A...465L..39H}   or micron-sized Fe-free silicates  \citep{Hoefner2008A&A...491L...1H, Norris2012Natur.484..220N} have been proposed as possible solutions to this dilemma. 

The high-spectral resolution of HIFI enables us to determine the expansion velocity of the spectral lines which sample different excitation regimes and hence are formed in different regions in the envelope. In addition, spectral lines (partially) formed in the wind acceleration region have a typical Gaussian profile \citep[instead of a flat-topped, two-horn or parabolic profile,][]{Lamers1999isw..book.....L}. Dedicated modelling can then yield some observational evidence on  the acceleration.


In the classical $\beta$-parametrization \citep[e.g.,][]{Lamers1999isw..book.....L} used to approximate the velocity structure, 
\begin{equation}
\varv(r)\simeq \varv_0+(\varv_\infty-\varv_0)\left(1-\frac{R_\star}{r}\right)^\beta\,,
\label{Eq:beta}
\end{equation}
with $\varv_0$ the velocity at the dust condensation radius and $\varv_\infty$  the terminal velocity, $\beta$ quantifies the wind acceleration. \citet{Schoier2011A&A...530A..83S} derived a value of $\beta=0.95$ for the S-type AGB star $\chi$ Cyg; for the oxygen-rich AGB star IK~Tau a value of 1 is derived \citep[see Fig.~\ref{FIG:IKTAU_VELOCITY} from][]{Decin2010A&A...521L...4D}. 
These are significantly more shallow velocity laws compared to that
derived by \citet{Schoier2006ApJ...649..965S} for the high-mass-loss rate
carbon-rich AGB star IRC\,+10216, where $\beta=0.2$. This illustrates that the wind driving and momentum coupling is indeed more efficient in carbon-rich AGB stars, but also that oxygen-rich and S-type AGB stars do reach their terminal velocity within $\sim$100\,\Rstar. 


\begin{figure}[htp]
 \includegraphics[width=\textwidth]{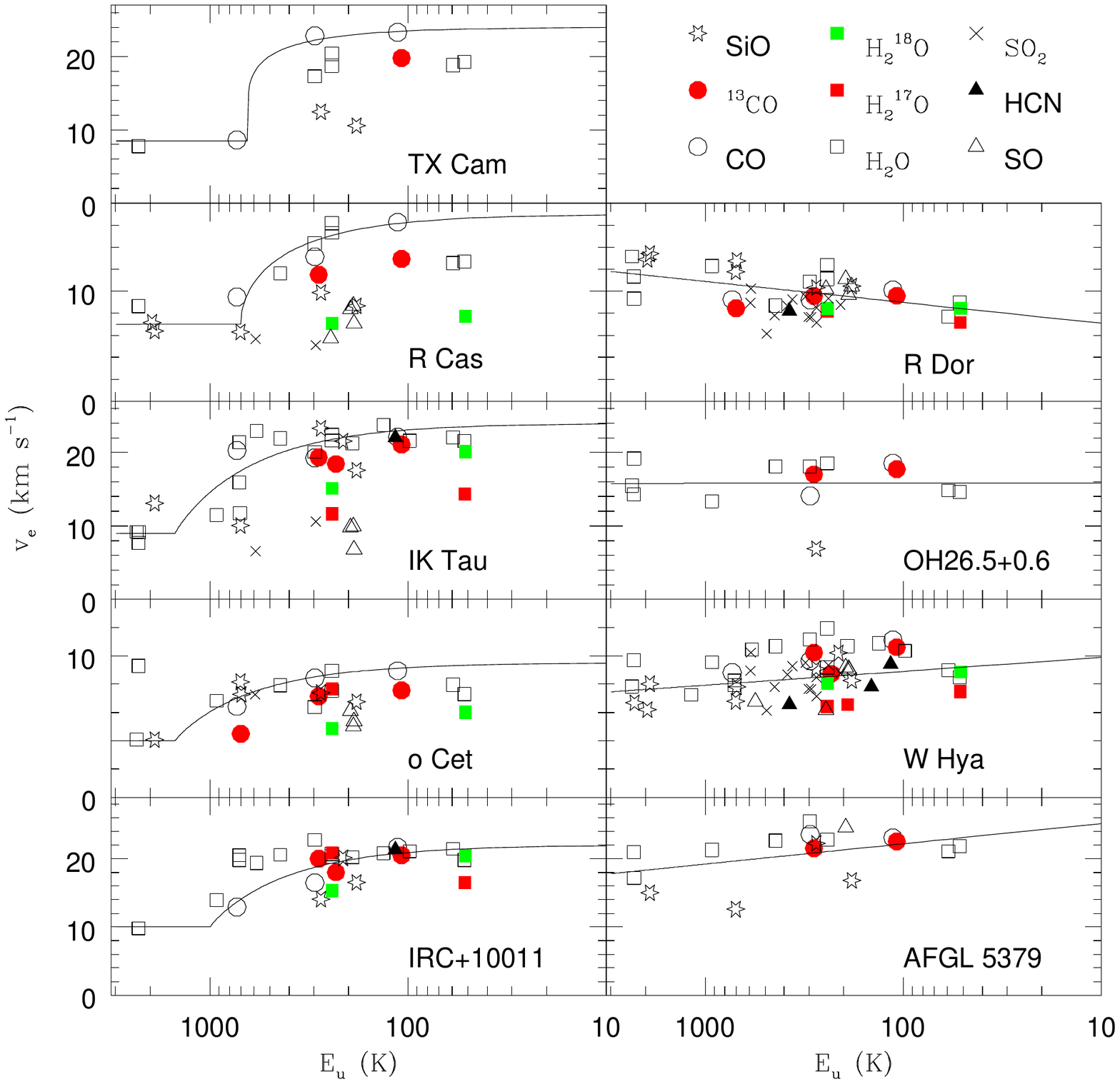}
\caption{Observed expansion velocity for different molecular
lines observed with \textit{Herschel}/HIFI in a sample of oxygen-rich AGB stars \citep{Justtanont2012A&A...537A.144J}.  The abscissa gives the upper energy level of the different lines, the ordinate the measured expansion velocity.}
\label{FIG:KAY_VELOCITY}
\end{figure}

An intriguing figure on observed expansion velocities in oxygen-rich AGB stars is presented by \citet{Justtanont2012A&A...537A.144J} (see Fig.~\ref{FIG:KAY_VELOCITY}). In that figure, a function similar to Eq.~\ref{Eq:beta} has been overplotted, but with a dependence on the energy rather than the radius:
\begin{equation}
\varv(r)\simeq \varv_0+(\varv_\infty-\varv_0)\left(1-\frac{E_0}{E}\right)^\beta\,,
\label{Eq:beta2}
\end{equation}
where $E_0$ is an arbitrary initial energy. Five targets show a clear wind acceleration: TX~Cam ($\beta=0.2$), R~Cas ($\beta=0.5$), IK~Tau ($\beta=0.9$), $o$~Cet ($\beta=0.9$), and IRC\,+10011 ($\beta=0.9$); in W~Hya and AFGL~5379 an increasing velocity w.r.t.\ the upper energy level is visible, although a strong onset of the acceleration is absent. In the case of R~Dor, however, the highly excited lines of
both H$_2$O and SiO indicate a larger expansion velocity in the inner
part of the envelope than for the lower J-transitions of CO. 

It is important to realize that the interpretation of this kind of figure  is not unequivocal since different molecules are involved, each having different optical depths and different line formation regions, and some lines are even masing. A varying turbulent velocity might play a role to explain the  velocity profiles showing an unexpected deceleration w.r.t.\ the upper energy level, but cannot explain the total spread.  In general, the turbulent velocity in the outer atmosphere and inner envelopes of  AGB stars is not well known. For non-Mira M-giants, the analysis of CO first-overtone bands indicates a turbulent velocity ranging from $\sim$1 to 4 km/s \citep{Tsuji1986A&A...156....8T}.  For the well-known carbon-rich AGB star IRC\,+10216, it was speculated that the turbulent velocity decreases somewhat with increasing distance from the star, from a value of a few km/s close to the stellar surface to $\sim$0.6\,km/s at hundreds of stellar radii \citep{Skinner1999MNRAS.302..293S}. Using interferometric 
observations, a turbulent velocity of (only) 1.5\,km/s in the inner wind of R~Dor was derived \citep{Schoier2004A&A...422..651S}, far less than the spread seen in Fig.~\ref{FIG:KAY_VELOCITY}. But also the high-resolution HIFI data offer the possibility to derive the turbulent velocity from a detailed analysis of the line wings. A recent example showing the strength of the HIFI data to estimate the turbulent velocity is shown in Fig.~8 in \citet{DeBeck2012}. This strength of the HIFI data is still largely unexplored.

It is clear that from a detailed analysis of the high-spectral resolution HIFI data, one can derive the global (1D) characteristics of the velocity field after a dedicated analysis of different lines with different excitation levels. However, line formation regions are usually quite broad (see Fig.~\ref{FIG:IKTAU_VELOCITY}), complicating a precise determination of the wind acceleration.

\subsection{CSM-ISM bow shock} \label{SECT:bow shock}

The \textit{Herschel} PACS and SPIRE images provide abundant scope for detailed studies on the complex interaction region between the circumstellar and interstellar material.

Photometric observations performed with the IRAS telescope \citep{Noriega1997AJ....114..837N}, and later confirmed with the Spitzer Space Telescope and AKARI \citep{Martin2007Natur.448..780M, Ueta2006ApJ...648L..39U, Ueta2008PASJ...60S.407U} indicated a new type of process that might impact  the chemical composition of circumstellar material (CSM) which is injected into the ISM: the interaction zone between the CSM and ISM might be more energetic than assumed earlier, resulting in a bow shock structure at the place where the wind collides with the ISM (in analogy with the solar environment). However, these early observations of the far-IR and UV-bright emission structures lack any spectral resolution, and as they have only poor spatial resolution, allow only basic morphological studies (e.g. Wilkin fitting \citep{Wilkin1996ApJ...459L..31W} to derive relative velocities).

\begin{figure*}[htp]
 \includegraphics[width=\textwidth]{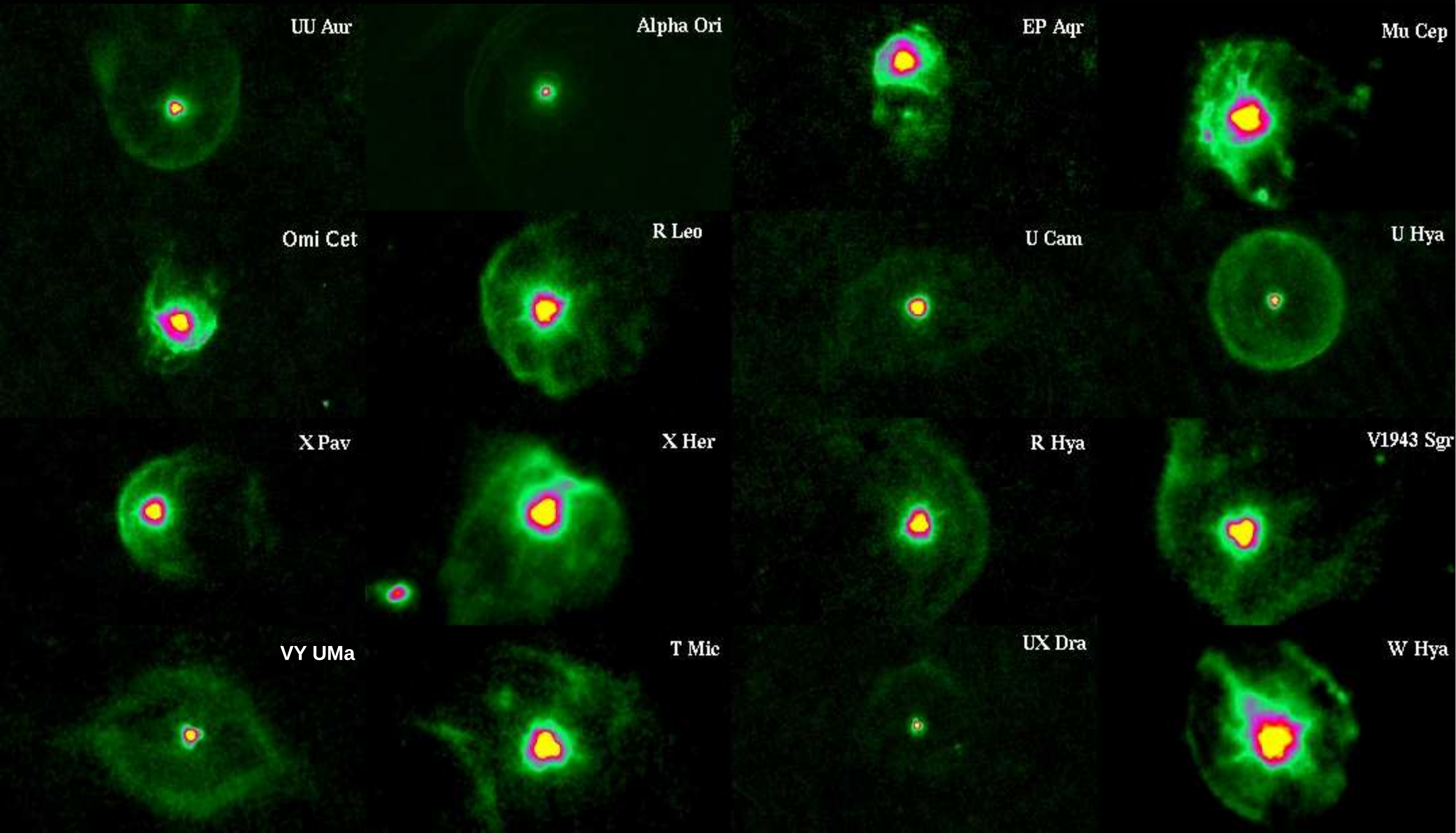}
\caption{PACS 70\,$\mu$m images of the CSM-ISM interaction of a sub-sample of evolved stars as observed in the MESS GTKP. More details and the image of other targets can be found in \citet{Cox2012A&A...537A..35C}. The images are best viewed on screen.}
\label{FIG:ISM_CSM}
\end{figure*}

In the MESS GTKP \citep{Groenewegen2011A&A...526A.162G} several known and new interaction objects have been imaged with the PACS instrument, revealing unprecedented detail. A nice overview is given by \citet{Cox2012A&A...537A..35C} (see also Fig.~\ref{FIG:ISM_CSM}). The detection rate of bow shocks is very high (80\%) when the subset is limited to distances less than 300\,pc! The detached shells around TT~Cyg, U~Ant, and AQ~And are discussed by \citet{Kerschbaum2010A&A...518L.140K} and on some more objects by \citet{Kerschbaum2011ASPC..445..589K}. Bow shocks are reported for CW~Leo \citep{Ladjal2010A&A...518L.141L}, X~Her and TX~Psc \citep{Jorissen2011A&A...532A.135J}, and $\alpha$ Ori \citep{Decin2012A}. For $o$~Cet the inner part of the stellar wind bubble bounded
and formed by the termination shock is seen \citep{Mayer2011A&A...531L...4M}.

\begin{figure}
 \includegraphics[width=\textwidth]{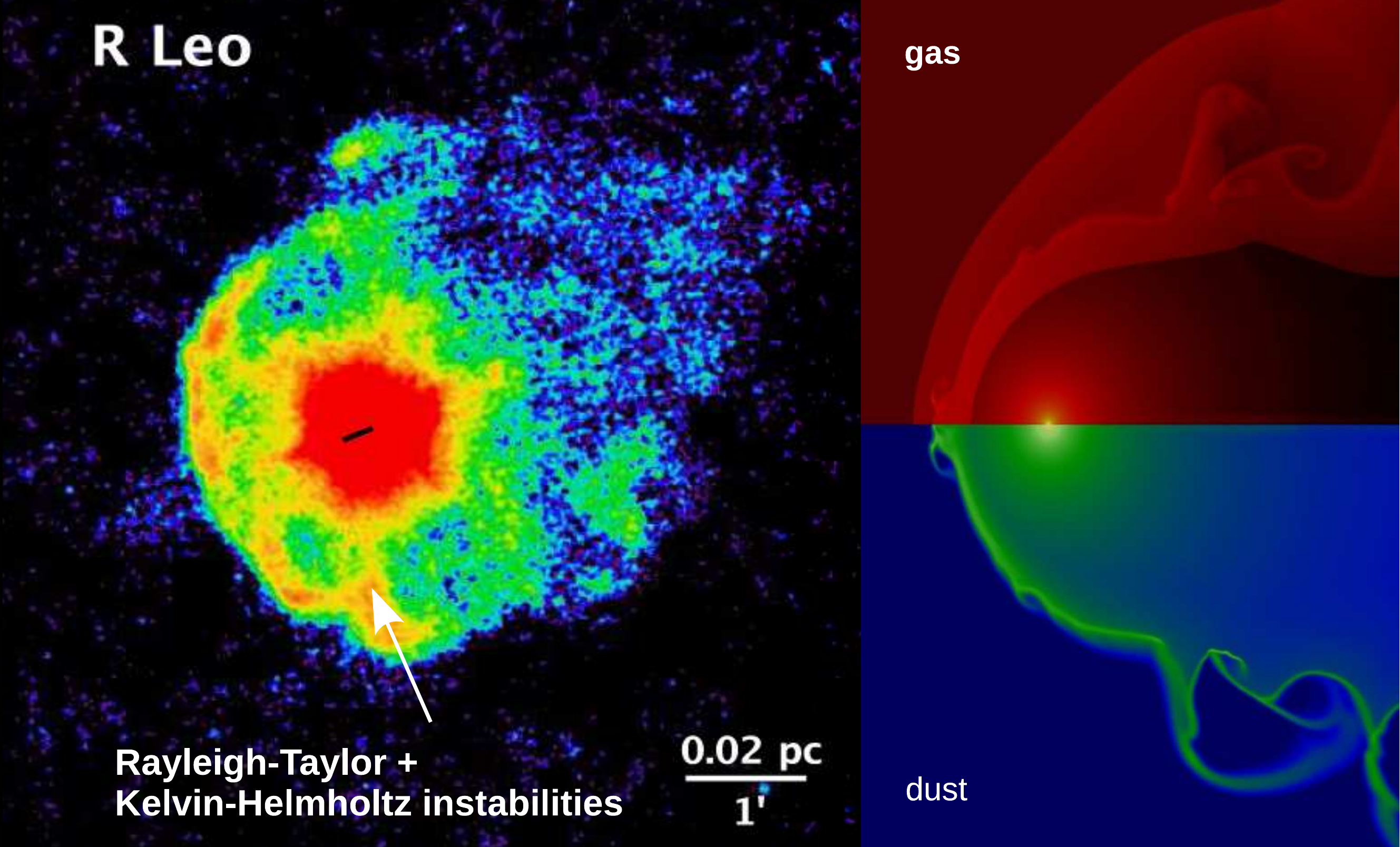}
\caption{PACS 70\,$\mu$m image of R~Leo (left) compared to a hydrodynamical simulation of the wind-ISM interaction computed with the {\tt AMRVAC} code \citep{vanMarle2011ApJ...734L..26V}. The parameters for this simulation are a stellar wind velocity of 15\,km/s, a constant gas mass-loss rate of $1 \times 10^{-6}$\,M$_\odot$/yr, a dust-to-gas mass ratio of 0.01, a space velocity of 25\,km/s, a local ISM density of 2 cm$^{-3}$, and an ISM temperature of 3\,K. The upper right figure shows the gas density, which range (in log-scale) between $10^{-24}$ and $10^{-19}$\,g/cm$^{3}$; the lower right figure represents the dust grain particle density, ranging (in log-scale) between $10^{-10}$ and $10^{-3.5}$\,cm$^{-3}$.}
\label{Fig:RLeo_hydro}
\end{figure}

The morphological classification introduced by \citet{Cox2012A&A...537A..35C} includes 'fermata's' (cfr.\ X~Pav), 'rings' (cfr.\ U~Hya), 'eyes' (cfr. VY~UMa), and 'irregular'-like structures. Depending on the space velocity, mass-loss rate, ISM temperature and density, etc.\ different bow shock structures might appear \citep[see Fig. 8--10 in][]{Cox2012A&A...537A..35C}, with clear signatures of Kelvin-Helmholtz\footnote{Kelvin-Helmholtz (KH) instabilities arise when velocity shear is present in a fluid or when there is sufficient velocity difference  across the interface between two fluids} and/or Rayleigh-Taylor\footnote{Rayleigh-Taylor (RT) instabilities might occur at the interface between two fluids of different densities,  when the lighter fluid is pushing the heavier fluid} instabilities. A textbook example in this respect is R~Leo (see Fig.~\ref{Fig:RLeo_hydro}), where clear `smaller scale' Rayleigh-Taylor  instabilities (visible in the form of mushroom-like signatures) and `large scale' Kelvin-
Helmholtz (deforming the RT instabilities along the contact discontinuity) are present. These \textit{Herschel} images are the first images showing both types of instabilities in the shocked wind-region around evolved AGB and supergiant stars.

\begin{figure*}
 \includegraphics[width=\textwidth]{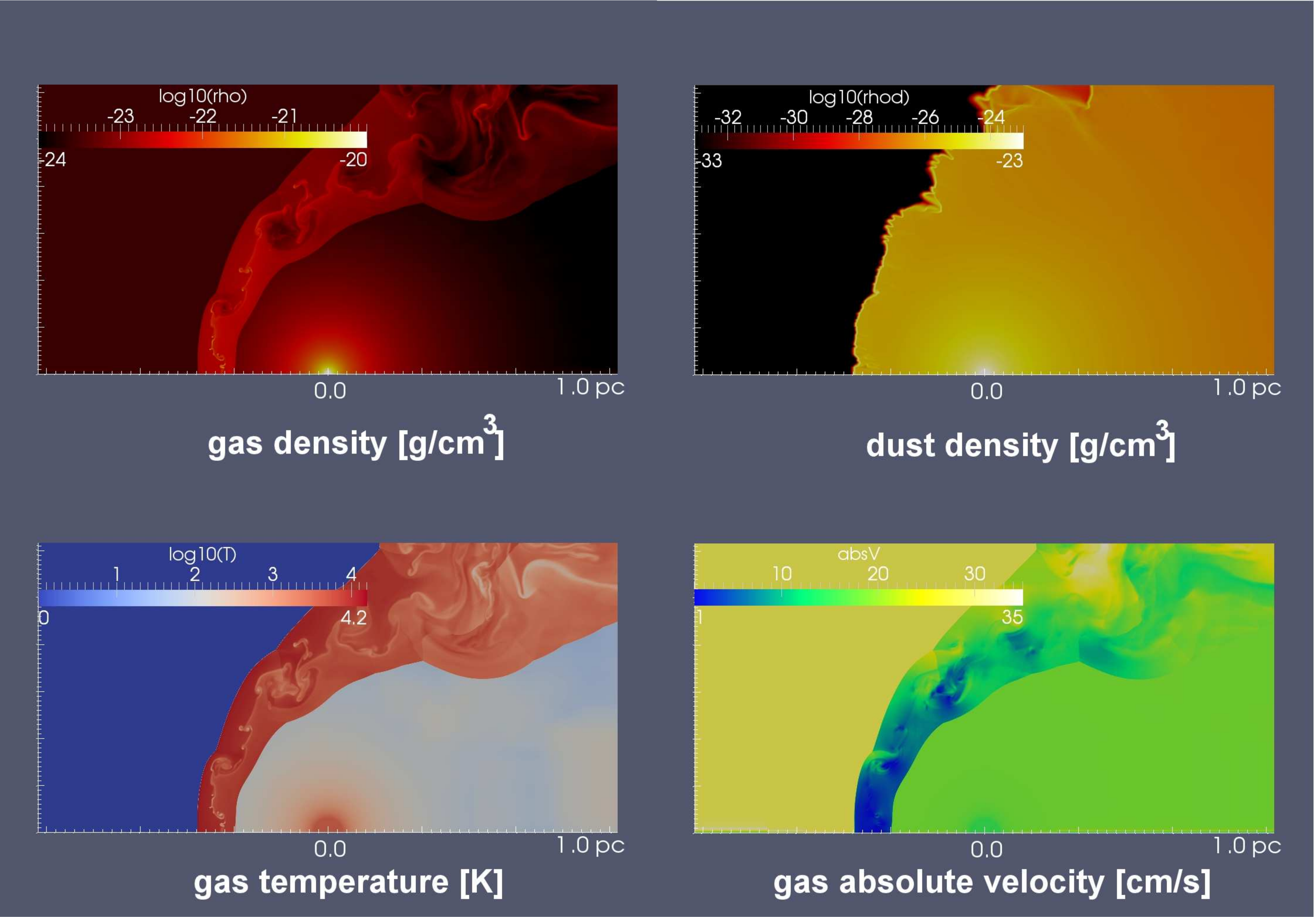}
\caption{Hydrodynamical simulation of the wind-ISM interaction of the envelope around Betelgeuse after 50\,000\,yr of evolution using the {\tt AMRVAC} code \citep{vanMarle2011ApJ...734L..26V}. The parameters for this simulation are a stellar temperature of 3650\,K, a stellar wind velocity of 15\,km/s, a constant gas mass-loss rate of $3 \times 10^{-6}$\,M$_\odot$/yr, a dust-to-gas mass ratio of $2.5 \times 10^{-3}$, a velocity  w.r.t.\ the local ISM of 28.3\,km/s, a local ISM density of $3.2 \times 10^{(-24)}$\,g/cm$^3$, and an ISM temperature of 3\,K. \textit{First panel:} gas density in g/cm$^3$, \textit{second panel:} dust density in g/cm$^3$ (for a dust particle with a radius of 0.05\,$\mu$m), \textit{third panel:} gas temperature in Kelvin, \textit{fourth panel:} gas absolute velocity in cm/s.
After a simulation time of 50\,000\,yr, the place of the bow\-shock interaction has stabilized:  the termination shock occurs at $\sim$0.33\,pc, the bow\-shock  at $\sim$0.46\,pc, and the turbulent  astropause/astrosheath in between. The instabilities are Rayleigh-Taylor instabilities (due to the density difference between the CSM and ISM material) which are somewhat 'dis-formed' due to the velocity shear between CSM and ISM (yielding Kelvin-Helmholtz instabilities). The gas temperature in the interaction zone can be as high as few thousand K, while the dust temperature only increases slightly w.r.t.\ the region without ISM interaction. Detailed inspection of the gas and dust density shows that a small dust particle (of 0.05\,$\mu$m in this simulation) is able to travel beyond the CSM-ISM contact discontinuity, but does not enter the unshocked ISM. A dust particle with a grain size of 1\,$\mu$m is able to cross the border between shocked and unshocked ISM material.}
\label{FIG:hydro}
\end{figure*}

\begin{figure*}
 \centering\includegraphics[width=.75\textwidth]{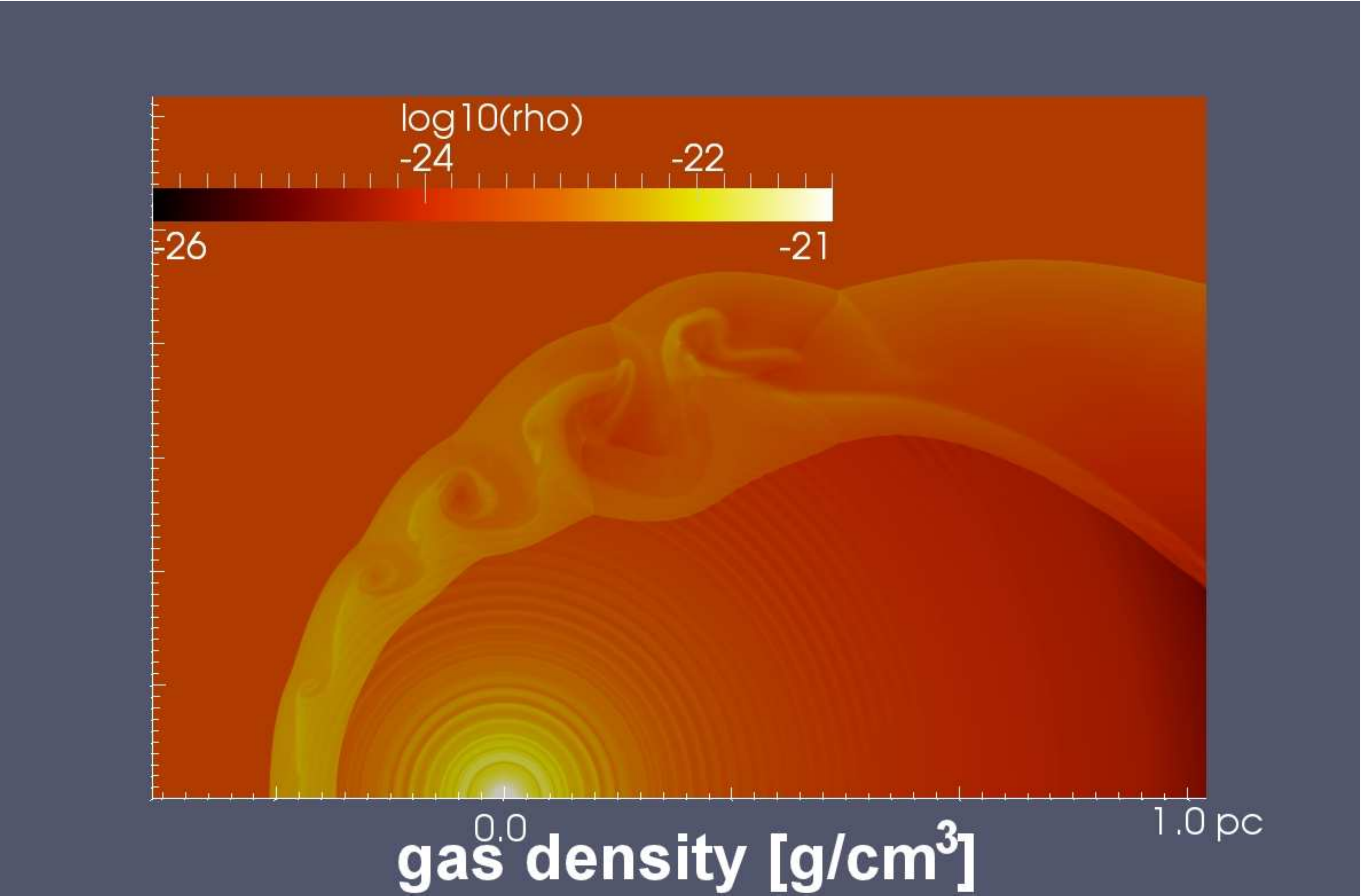}
\caption{Same hydrodynamical simulation as in Fig.~\ref{FIG:hydro}, but for a varying mass-loss rate with a period of 1000\,yr, implemented as \Mdot($t$)\,=\,$1.8 \times 10^{-6} + (1.5 \times 10^{-6} \times \sin(2 \pi t/1000\,{\rm yr}))$\,\Msun/yr.}
\label{FIG:hydro_varmot}
\end{figure*}

The new \textit{Herschel} data trigger the development of hydrodynamical codes to simulate these environments \citep[e.g.][]{vanMarle2011ApJ...734L..26V, Mohamed2012A&A...541A...1M}.
Fig.~\ref{FIG:hydro} shows the wind-ISM interaction region obtained 
from a hydrodynamical simulation using the  {\tt AMRVAC} code \citep{Keppensetal:2012}. This code has recently been improved \citep{vanMarle2011ApJ...734L..26V} to include dust grains in the stellar wind,
and to take the drag forces between dust and gas into account. The example shows that the dust is only heated up by a few K in the interaction zone, while the gas temperature can be as high as 10\,000\,K. Clumps in the inner envelope (as seen, e.g., in the \textit{Herschel} images described in Sect.~\ref{SECT:GEO_INNNER}) might have an imprint on the bow shock morphology. 
Simulations show that a smoothly varying (sinusoidal) wind density, as shown in Fig.~\ref{FIG:hydro_varmot}, has an imprint in the free flowing wind, but that the density contrast is completed erased at the turbulent contact discontinuity. Only very strong 'picket-fence'-like variations with a factor of a few thousand might impact the bow shock morphology \citep{Decin2012A}.

The `ring'-like bow shocks (as U~Hya) \textit{all} belong to thermally-pulsating stars, being carbon-rich AGB stars or technetium enriched M-type supergiants. The origin of the detached shells appears linked to the interaction of a fast wind with an 'older' slow wind or a dramatic change in mass loss caused by, e.g., a thermal pulse (TP). 

\begin{figure}[htp]
 \centering{\includegraphics[width=.75\textwidth]{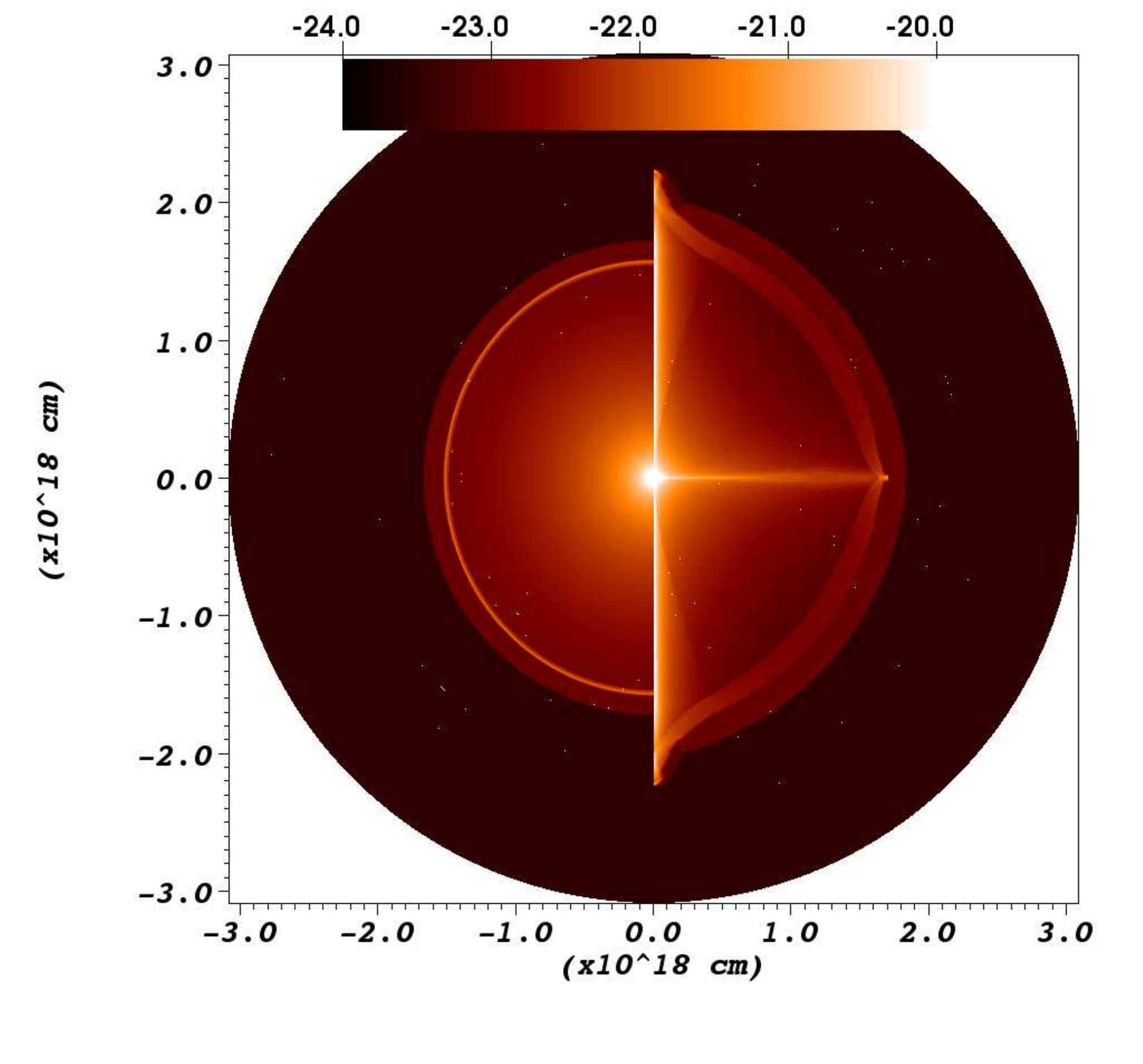}}
\caption{Logarithm of the density in g cm$^{-3}$ for the expansion of an AGB wind into a constant density ISM; one without magnetic field ($\sigma$\,=\,0, \textit{left}) and one with $B=0.931$\,G (for $\sigma$\,=\,0.1, \textit{right}). The non-magnetic simulation shows a very thin shocked wind layer and a thicker shocked ISM shell. 
For the simulation with magnetic field, both layers are approximately equally thick. The magnetic field also creates collimated streams of matter over the poles and at the equator, 
which distort the spherical symmetry of the shell.}
 \label{AGB_highB}
\end{figure}

The structures least understood, in my opinion, are the 'eye'-like structures (as VY UMa),  two elliptical non-concentric arcs observed at opposing sides of the central source with a covering angle of $\le$180\deg. For a few sources,  the two arcs are connected and there is even tentative evidence for a jet structure in the mid plane. Binarity might play a role, but this can not be confirmed or excluded at the moment. Magnetic fields might play a role as well, but this field is still totally unexplored. As an example, we show in Fig.~\ref{AGB_highB} a simulation using the {\tt MPI-AMRVAC} code 
\citep{Keppensetal:2012} for a star, on which a toroidal magnetic field 
\begin{eqnarray}
 B_\phi~=~B_\star\left(\frac{v_{\rm{rot}}}{v_\infty}\right)\left(\frac{R_\star}{r}\right)^2 \left(\frac{r}{R_\star}-1\right) \sin{(\theta)} \left(1-\frac{2\theta}{\pi}\right)
\label{eq:B1}
\end{eqnarray}
(with $v_{\rm{rot}}$ the surface rotation of the star, $R_\star$ the stellar radius, $r$ the radial grid coordinate, and $\theta$ the colatitudinal angle) has been imposed.
The stellar magnetic field $B_\star$ is defined as a fraction of the kinetic energy of the wind \citep{BegelmanLi:1992}:
\begin{eqnarray}
 \sigma~=~\frac{B_\star^2 R_\star^2}{\Mdot v_\infty} \left(\frac{v_{\rm{rot}}}{v_\infty}\right)^2.
\label{eq:B2}
\end{eqnarray}
This setup for the magnetic field is similar to the one used by \citet{GarciaSeguraetal:1999}, 
except that  Eq.~\ref{eq:B1} reverses the polarity of the magnetic field at the equator. 
The stellar parameters are $R_\star=150\Rsun$ and $v_{\rm{rot}}=10$\,km/s. The influence of rotation on the stellar wind was not included. The simulation was run for $\sigma=0$ (no magnetic field) and  $\sigma=0.1$ (yielding a stellar magnetic field $B_\star=0.931$\,Gauss)\footnote{The detected longitudinal field in the red supergiant Betelgeuse is $\sim$1\,Gauss \citep{Auriere2010A&A...516L...2A}}. 
This simulation shows indeed an 'eye'-like shape, but we should take into account that no magnetic fields are yet detected in AGB stars.

I am convinced that these \textit{Herschel} images are opening a new era of detailed studies of this intriguing interaction zone between ISM and CSM. Highly sophisticated model simulations and dedicated observations (with, e.g., ALMA) will elucidate the key astrophysical quantities triggering the appearance of  this whole zoo of interaction phases.

\section{New chemical insights from \textit{Herschel}} \label{SEC:CHEMISTRY}

\begin{figure*}[!htp]
\includegraphics[width=\textwidth]{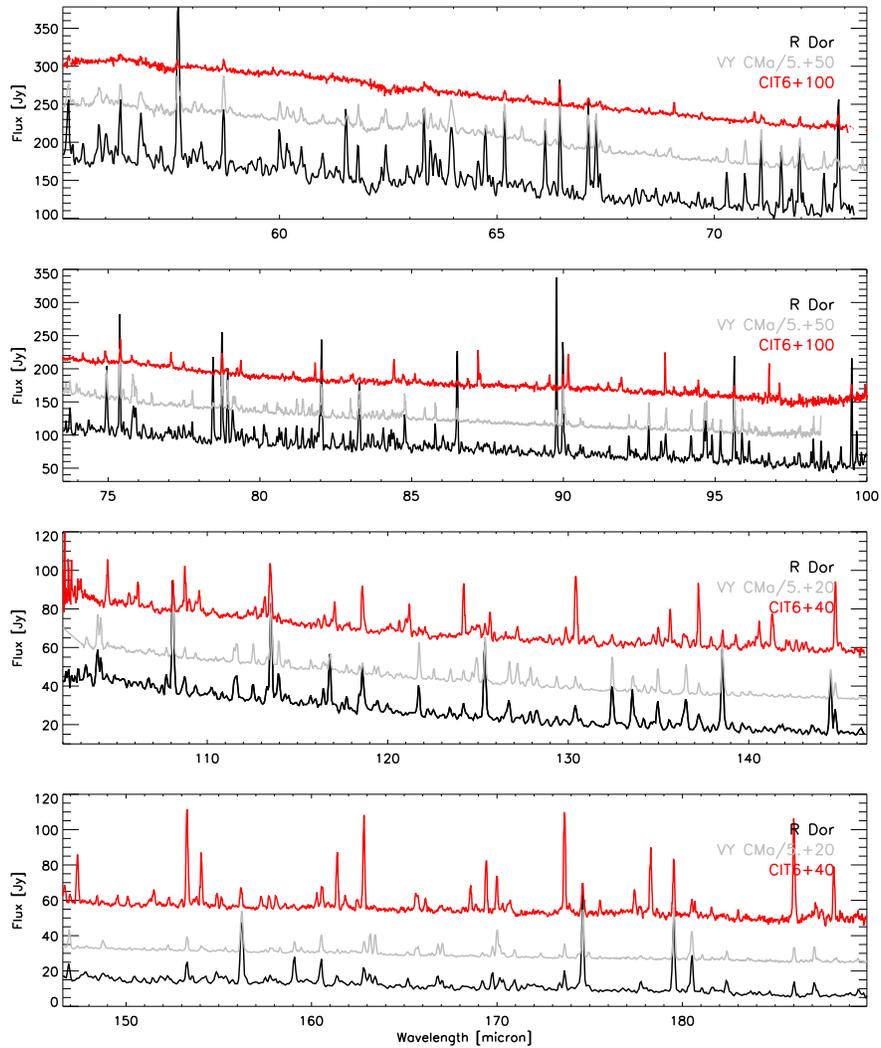}
\vspace*{1ex}
\caption{Comparison between the rich \textit{Herschel}/PACS spectra of the O-rich AGB star R~Dor (black), the O-rich supergiant VY~CMa (grey) and the C-rich AGB star CIT~6 (red). Spectra of VY~CMa and CIT~6 are scaled to facilitate the comparison. Most lines are blends, with he most prominent features being  CO and H$_2$O transitions (O-rich) and HCN (C-rich). Each spectrum contains some few hundred lines. Line identifications can be found in \citet{Decin2010Natur.467...64D, Decin2010A&A...518L.143D, Royer2010A&A...518L.145R}. }
\label{FIG:PACS_RDOR_VYCMA_CIT6}
\end{figure*}

To date, more than 70 molecules have been detected in the circumstellar envelopes (CSEs) of evolved stars \citep[e.g.][]{Cernicharo2000A&AS..142..181C, He2008ApJS..177..275H}. Admittedly, most of them are only detected in the nearest carbon-rich AGB star IRC\,+10216. The effect of nucleosynthesis in the stellar core and subsequent convective envelope mixing or dredge-ups determine the abundance structure of the stellar outer atmospheric layers. Thermodynamic equilibrium and non-equilibrium reactions, photochemical reactions, ion-molecule reactions, and the condensation of dust grains establish the abundance stratifications throughout the circumstellar envelope (see Fig.~\ref{FIG:scheme_AGB}). The IRAS, Spitzer and AKARI data, and recently the Herschel PACS and SPIRE images (see Sect.~\ref{SECT:bow shock}) indicate that the CSM-ISM interaction region might be a chemically complex region. 

The three spectrometers on board \textit{Herschel} have proven to be three powerful instruments to unravel these complex chemical characteristics. Some first results are given in the subsequent sub-sections covering gradually the full envelope starting at the innermost regions. It will become clear that most of these first results are based on thorough non-local thermodynamic equilibrium (non-LTE) calculations, sometimes invoking the modeling of different molecules at once. Indeed, in the era before \textit{Herschel}, often only a handful of lines (at maximum) of one specific molecule were modeled by solving the statistical equilibrium rate equations. Today, often hundred molecular line transitions of several molecules are simultaneously modeled (see Fig.~\ref{FIG:PACS_RDOR_VYCMA_CIT6}), if needed by using supercomputers. In addition, it becomes more and more clear that a consistent dust-gas modeling effort is of utmost importance to correctly derive the physical and chemical envelope structure \citep[see, e.
g.,][]{Lombaert2012}.

\subsection{Nucleosynthesis and dredge-ups} \label{SECT:nucleo_du}

The material injected in the ISM reflects the effects of nucleosynthesis in the stellar core and subsequent convective envelope mixing or dredge-ups (DUs). The effect of nucleosynthesis and DUs on the stellar abundances can be constrained by different observations: one can study the abundance pattern (1) in the dense outer layers of the stellar AGB atmosphere, (2) in the tenuous extended CSE, and (3) from presolar grains which are thought to be condensed from AGB stars. However, stellar atmosphere analysis excludes high mass-loss AGB stars and presolar grains have unpredictable selection effects. Due to these observational selection effects, the injected abundance pattern is best studied in the stellar wind of the AGB or supergiant star. 

In the cool CSE,  several chemical processes (see Fig.~\ref{FIG:scheme_AGB}) can alter the molecular abundance fractions, originally established in the dense stellar atmosphere (as a result of the nucleosynthesis and DU events). Hence, the absolute abundance of most molecules can not be used as direct tracer of nucleosynthesis and subsequent DUs\footnote{Some indirect hints can, however, be derived. E.g., the low HF-abundance derived for IRC\,10216 argues towards the fact that $^{19}$F nucleosynthesis is not effective in AGB stars \citep{Agundez2011A&A...533L...6A}.}.  \textit{Isotopic} ratios are by far the most diagnostic tracers of the stellar origin of elements, as they are very sensitive to the precise conditions (especially the temperature) in the nuclear burning regions, and different isotopes of the same element can form in very different reaction chains \citep[e.g.][]{Karakas2008ApJ...676.1254K}. Moreover, comparisons of isotopologue fractions are less affected by systematic effects than are 
comparisons of the abundances of different molecules. There is, however, 1 chemical process which might affect some isotopologue species differently (as CO and CS): chemical fractionation \citep{Mamon1988ApJ...328..797M}. CO chemical fractionation could enhance the abundance of $^{13}$CO in the outer envelope. However, chemical fractionation competes with selective photodissociation, a process that preferentially destroys $^{13}$CO but does not significantly affect optically-thick $^{12}$CO \citep{Mamon1988ApJ...328..797M}. These two mechanisms are thought to usually compensate each other \citep{Milam2009ApJ...690..837M}.

However, observed isotope fractions for AGB and supergiant winds are (still) very scarce.
Although, equipped with a relative small mirror compared to some ground-based facilities, \textit{Herschel} is sensitive enough to observe some of the less abundant isotopologue species. Line emission of ortho- and para-water in the different isotopologue states H$_2^{16}$O, H$_2^{17}$O, and H$_2^{18}$O, as well as $^{12}$C$^{16}$O, $^{13}$C$^{16}$O, $^{12}$C$^{17}$O, $^{12}$C$^{18}$O, $^{28}$Si$^{16}$O, $^{29}$Si$^{16}$O, $^{30}$Si$^{16}$O, H$^{35}$Cl, H$^{37}$Cl, $^{28}$Si$^{32}$S, $^{29}$Si$^{32}$S, $^{30}$Si$^{32}$S, $^{28}$Si$^{33}$S, $^{28}$Si$^{34}$S, $^{12}$C$^{32}$S, $^{13}$C$^{32}$S, $^{12}$C$^{33}$S, $^{12}$C$^{34}$S, H$^{12}$C$^{14}$N,  H$^{13}$C$^{14}$N etc.\ are detected in the HIFI, PACS and SPIRE spectra \citep{Royer2010A&A...518L.145R, Decin2010A&A...521L...4D, Decin2010A&A...518L.143D, Schoier2011A&A...530A..83S, Agundez2011A&A...533L...6A}. Often, one can only deduce a lower limit to the isotopologue fraction\footnote{defined as the fraction of the most abundant species to the least 
abundance species} due to opacity effects. Analysing several lines covering different excitation regimes partly solves this issue.

For the red supergiant VY~CMa \citet{Royer2010A&A...518L.145R} derived an ortho-to-para ratio (OPR) for H$_2$O  of 1.27, significantly lower than the expected value of 3$^($\footnote{reflecting a high-temperature ($\sim$50) equilibrium water formation}$^)$. If opacity effects are not compromising these results, this reflects a spin temperature $\sim$15\,K indicating formation of H$_2$O in non-LTE conditions.
\citet{Schoier2011A&A...530A..83S} derived for the S-type AGB star $\chi$ Cyg an ortho-to-para OPR H$_2$O ratio of $1.4\pm0.34$, a $^{12}$CO/$^{13}$CO ratio of $43 \pm 6$, a H$^{12}$CN/H$^{13}$CN ratio of $56\pm28$, a $^{28}$SiO/$^{29}$SiO ratio of $8 \pm 2$, and a $^{28}$SiO/$^{30}$SiO ratio of $19 \pm 8$. The quite high $^{12}$C/$^{13}$C ratio suggests that this ratio is higher in S-type than in M-type AGB stars, and more in  line with those derived for carbon stars. The Si isotopic ratios can be compared with the $^{28}$Si/$^{29}$Si and $^{28}$Si/$^{30}$Si ratios inferred for the Sun (19 and 29, respectively), the M-type AGB star IK~Tau \citep[13 and 40, respectively,][]{Decin2010A&A...521L...4D}, and IRC\,+10216 \citep[15
and 20, respectively,][]{Kahane2000A&A...357..669K}. Again, these isotopic ratios of $\chi$~Cyg are more in line with the carbon-rich AGB star IRC\,+10216.

For another AGB target, the O-rich AGB star IK~Tau, \citet{Decin2010A&A...521L...4D} derived in OPR for H$_2$O of 3, and H$_2^{16}$O/H$_2^{17}$O\,=\,$600\pm150$, H$_2^{16}$O/H$_2^{18}$O\,=\,$200\pm50$ (well below the solar values of $^{16}$O/$^{17}$O$\sim$2632 and $^{16}$O/$^{18}$O$\sim$499). The lower than solar $^{16}$O/$^{18}$O ratios cannot be explained by any stellar evolution model in the literature. However, IK~Tau is not the only Galactic star with a low $^{16}$O/$^{18}$O ratio: some barium stars analyzed by \citet{Harris1985ApJ...292..620H} also have a low $^{16}$O/$^{18}$O value. The observation of the higher excitation $^{28}$SiO J=14-13 line with HIFI, resulted in a reduction of the $^{28}$SiO abundance in the inner wind with a factor 2 compared to the value derived by \citet{Decin2010A&A...516A..69D}, proving the strength of observing higher-excitation lines.

It is anticipated that the \textit{Herschel} observations of other evolved stars, complemented with other ground-based data sets,  will add new information to this discussion. If not too much plagued by opacity effects, these will yield useful constraints for stellar evolution models.

\subsection{Non-equilibrium chemistry in the inner envelope}\label{SECT:non-TE}

 For a long time, the gas chemical composition in the CSE was believed to be dominated entirely by the C/O ratio of the photosphere. A C/O ratio greater than one implied that all the
oxygen was tied in CO, leading to an oxygen-free chemistry (the
so-called carbon stars), whereas a C/O ratio of less than one meant
that no carbon bearing molecules apart from CO could ever form in an
oxygen-rich (O-rich) environment (M-type stars). Stars having a
C/O-ratio $\sim 1$ are called S-type AGB stars. This picture, based
essentially on thermal equilibrium considerations \citep{Tsuji1973A&A....23..411T} applied to the gas, has been disproved by the detection of SiO at
millimeter (mm) wavelength in carbon-rich (C-rich) AGBs \citep[e.g.][]{Bujarrabal1994A&A...285..247B}. As for O-rich AGBs,
CO$_2$ infrared (IR) transition lines were detected in various objects
with the Short-Wavelength Spectrometer (SWS) on board the Infrared
Space Observatory (ISO) \citep[e.g.][]{Justtanont1998A&A...330L..17J}.

Originally, it was  thought that the carbon species observed in the winds of O-rich stars were produced via photochemical processes in the outer envelope.
However, \citet{Duari1999A&A...341L..47D} showed that pulsation-driven shock-induced
non-equilibrium chemistry models predict the formation of large
amounts of a few carbon species, like HCN, CS and CO$_2$, in the
inner envelope of O-rich AGB stars: these molecules are formed in
the post-shocked layers and are then ejected in the outer wind as
`parent' species. Later on, it was shown theoretically \citep{Cherchneff2006A&A...456.1001C} and observationally \citep{Decin2008A&A...480..431D}  that the inner winds of AGB stars show a striking homogeneity, independent of their C/O ratio and stage of evolution.

The \textit{Herschel} observations have confirmed this picture, showing that integrated line intensities ratios are often much larger than expected from thermodynamic equilibrium (TE) chemical production models \citep[e.g.][]{Royer2010A&A...518L.145R, Decin2010A&A...518L.143D, Schoier2011A&A...530A..83S}. The derived inner wind abundance might be orders of magnitude higher than the TE predictions. 

\begin{figure}
\centering \includegraphics[width=\textwidth]{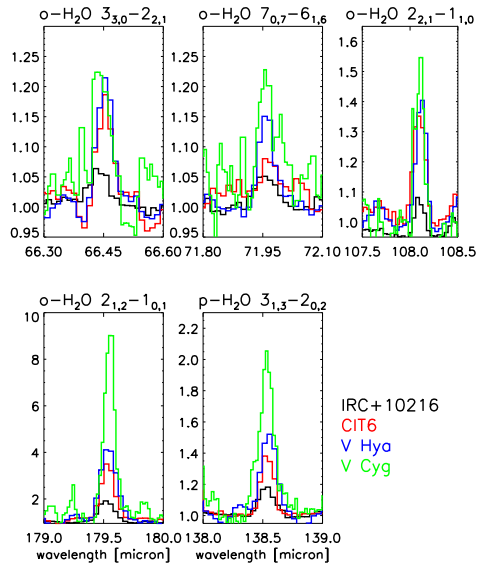}
\caption{Relative line strength of some water vapor lines as observed with \textit{Herschel}/PACS in the bright carbon sources IRC\,+10216, V~Hya, CIT~6, and V~Cyg. The line-to-continuum ratio is clearly highest for V~Cyg and lowest for IRC\,+10216, and can vary with a factor of a few from source to source.}
\label{FIG:C_stars_H2O}
\end{figure}

However, while the observations are showing overwhelming evidence for the occurrence of complex non-equilibrium chemistry, a direct confrontation with theoretical model predictions is still lacking. The reason for this is that the non-equilibrium chemistry models as presented by, e.g., \citet{Agundez2006ApJ...650..374A} and \citet{Cherchneff2006A&A...456.1001C}
only include a very restricted set of (circum)stellar parameters tuned toward predictions in agreement with the observations for one specific target. This was also the concept in the recent publication of \citet{Cherchneff2011A&A...526L..11C}. To explain the recent detection of warm water vapor in the inner envelope of the C-rich AGB star IRC\,+10216 by \citet{Decin2010Natur.467...64D}, the chemical model was updated by including a greater completeness of the chemistry, and including a more accurate chemistry for Si and S. The impact of this updated chemical network was huge, with the water vapour abundance at $\sim$3\,\Rstar\ increasing with few orders of magnitude.  
However, it is difficult to translate this type of results to other evolved stars without proper knowledge of the sensitivity of the results to different physical, chemical and numerical input parameters. Indeed, water vapor is now detected in all carbon-rich AGB stars surveyed with \textit{Herschel}, with the exception of 1 target. An analysis of the water-rich sources shows a large variation in the derived water abundances (see Fig.~\ref{FIG:C_stars_H2O}). E.g., V~Hya (with a mass-loss rate, \Mdot, of $1 \times 10^{-6}$\,M$_\odot$/yr) and CIT~6 (\Mdot=$4.2 \times 10^{-6}$\,M$_\odot$/yr) have a water peak abundance relative to H$_2$, $f_ {\rm{H_2O}}$, of $4.5 \times 10^{-6}$ and $3 \times 10^{-5}$, respectively. This is up to 2 orders of magnitude larger than in the case of IRC\,+10216 (\Mdot=$2.1 \times 10^{-5}$\,M$_\odot$/yr, $f_{\rm{H_2O}}$=$2\times 10^{-7}$), and a factor of 10--100 larger than the predictions of \citet{Agundez2010ApJ...724L.133A}. Currently, we do not know the key parameter(s) 
determining the water abundance.

\subsection{Dust mineralogy and gas-grain reactions} \label{SECT:dust_gas}

\subsubsection{Dust mineralogy} \label{SECT:dust_mineralogy}
The spectra of both AGB stars and supergiants are ideal tracers of the solid-state and molecular species that form in the cool CSE (see Fig.~\ref{FIG:PACS_RDOR_VYCMA_CIT6}). Most of the astronomical solid state features are found in the near and mid-IR ranges. ISO, especially with its SWS and LWS revolutionized our knowledge of dust and ice around stars \citep{Waters2004ASPC..309..229W}. In the LWS range, overlapping with \textit{Herschel} PACS, most of ISO's spectroscopic dust observations were suffering from signal-to-noise problems in all but the brightest AGB stars. The higher sensitivity of \textit{Herschel} could be crucial in identifying/analyzing few of these solid-state features. Dust-species like crystalline water-ice at 61\,$\mu$m, forsterite at 69\,$\mu$m,  hibonite at 78\,$\mu$m, and calcite at 92.6\,$\mu$m are expected or have already been detected by ISO \citep{Molster1999A&A...350..163M, Molster2002A&A...382..222M, Kemper2002A&A...394..679K, Mutschke2002A&A...392.1047M}. Other measured 
features lack an identification, e.g., the 62--63\,$\mu$m feature with candidate substances like dolomite, ankerite, or diopside \citep{Koike2000A&A...363.1115K, Kemper2002A&A...394..679K}. Polycyclic aromatic hydrocarbon (PAH) features have been predicted to occur at far-IR wavelengths \citep{Joblin2002MolPh.100.3595J}. The detection of crystalline silicates and water ice features in the PACS range is an indicator of the formation of very large oxygen-rich dust grains in the cool circumstellar shells around some intermediate-mass transition sources evolving from the AGB to the PN stage. Alternatively, these oxygen-rich dust grains surrounded by water ice mantles may  form in the coolest regions of long-lived circumstellar disks around (binary?) low-mass stars, which can then be preserved from destruction by the increasing UV field generated by the central star as it evolves towards the PN stage.

\begin{figure}[htp]
 \begin{minipage}[t]{.48\textwidth}
        \centering{\includegraphics[width=\textwidth]{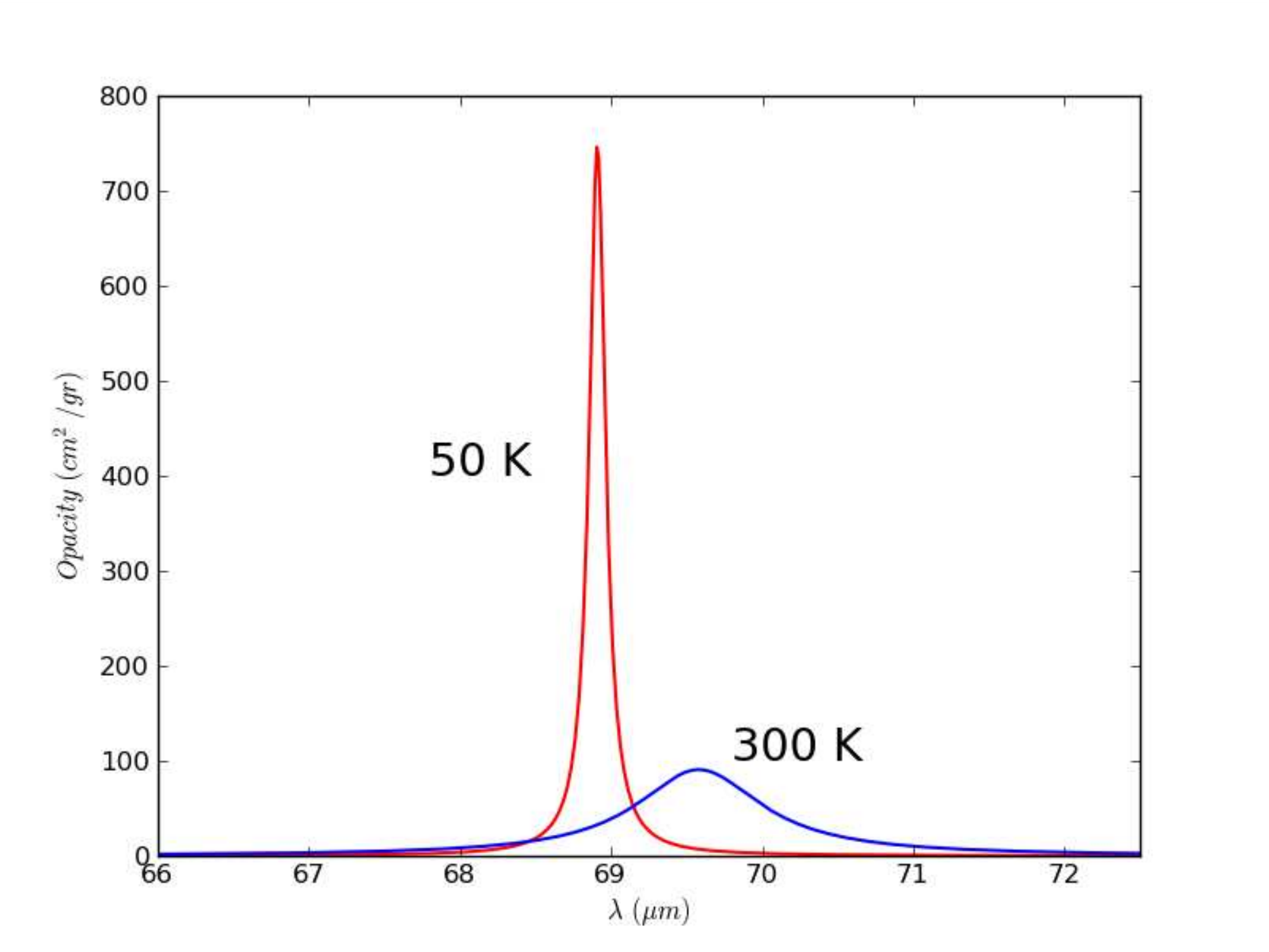}}
    \end{minipage}
 \begin{minipage}[t]{.48\textwidth}
        \centering{\includegraphics[width=\textwidth]{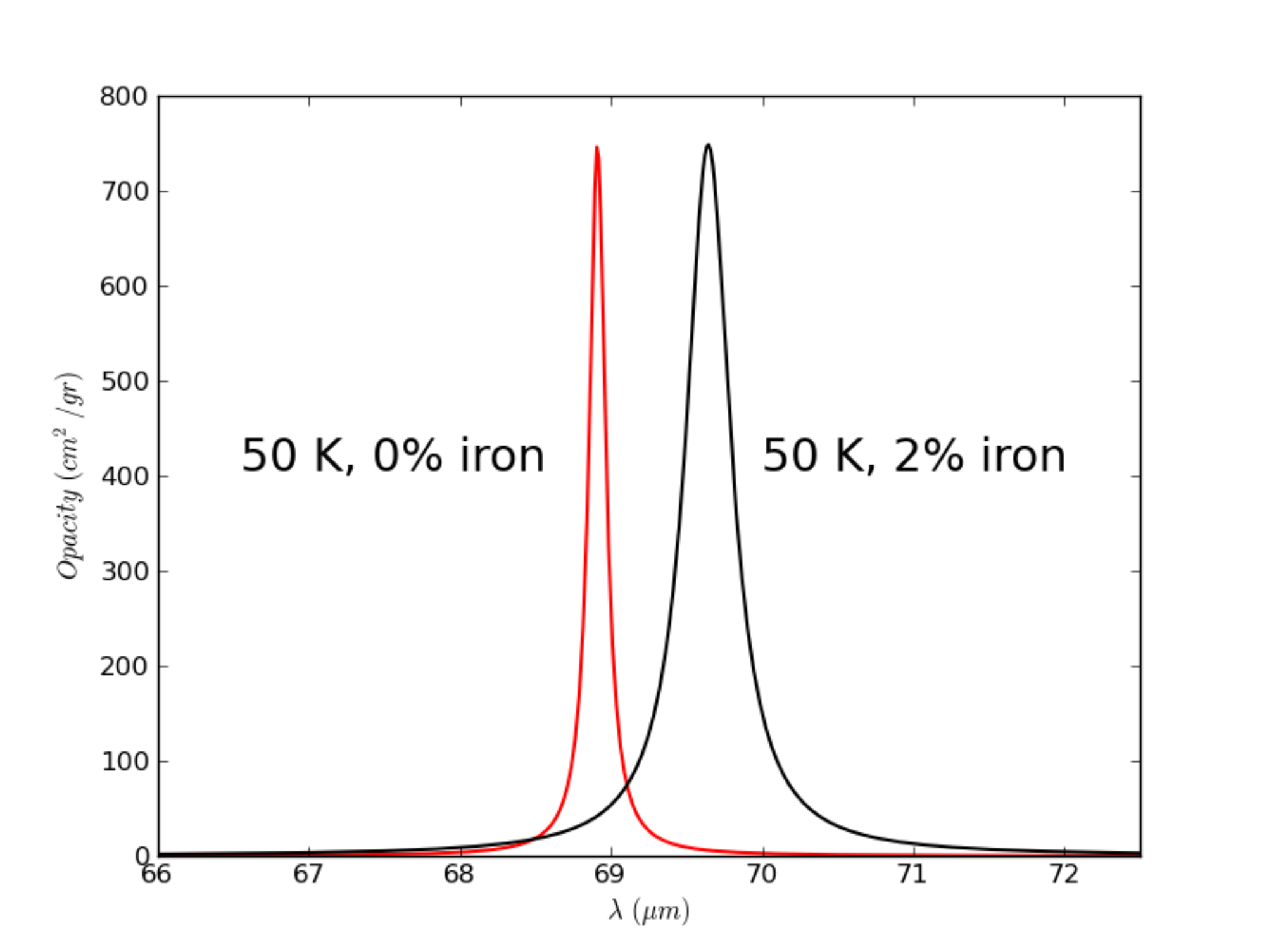}}
    \end{minipage}
\caption{\textit{Left-hand figure:} The 69\,$\mu$m band of forsterite (Mg$_2$SiO$_4$) is of special interest because its opacity is sensitive to the temperature of the grains. If the grains become hotter, the band shifts to the red, becomes broader and it drops in  strength \citep{Suto2006MNRAS.370.1599S}. This is illustrated in case of a temperature of 50 (red) and 300\,K (blue). 
\textit{Right-hand figure:} Forsterite is a crystalline olivine ((Mg, Fe)$_2$SiO$_4$) with a Mg-Fe composition that is purely magnesium rich.  The opacity is dependent on the iron content \citep{Koike2006A&A...449..583K}: starting with pure forsterite and increasing the iron content of the crystalline olivine, the 69\,$\mu$m band shifts to the red and becomes slighty broader.}
\label{Fig:Ben_didactiek}
\end{figure}

\begin{figure}[htp]
        \centering{\includegraphics[width=.75\textwidth]{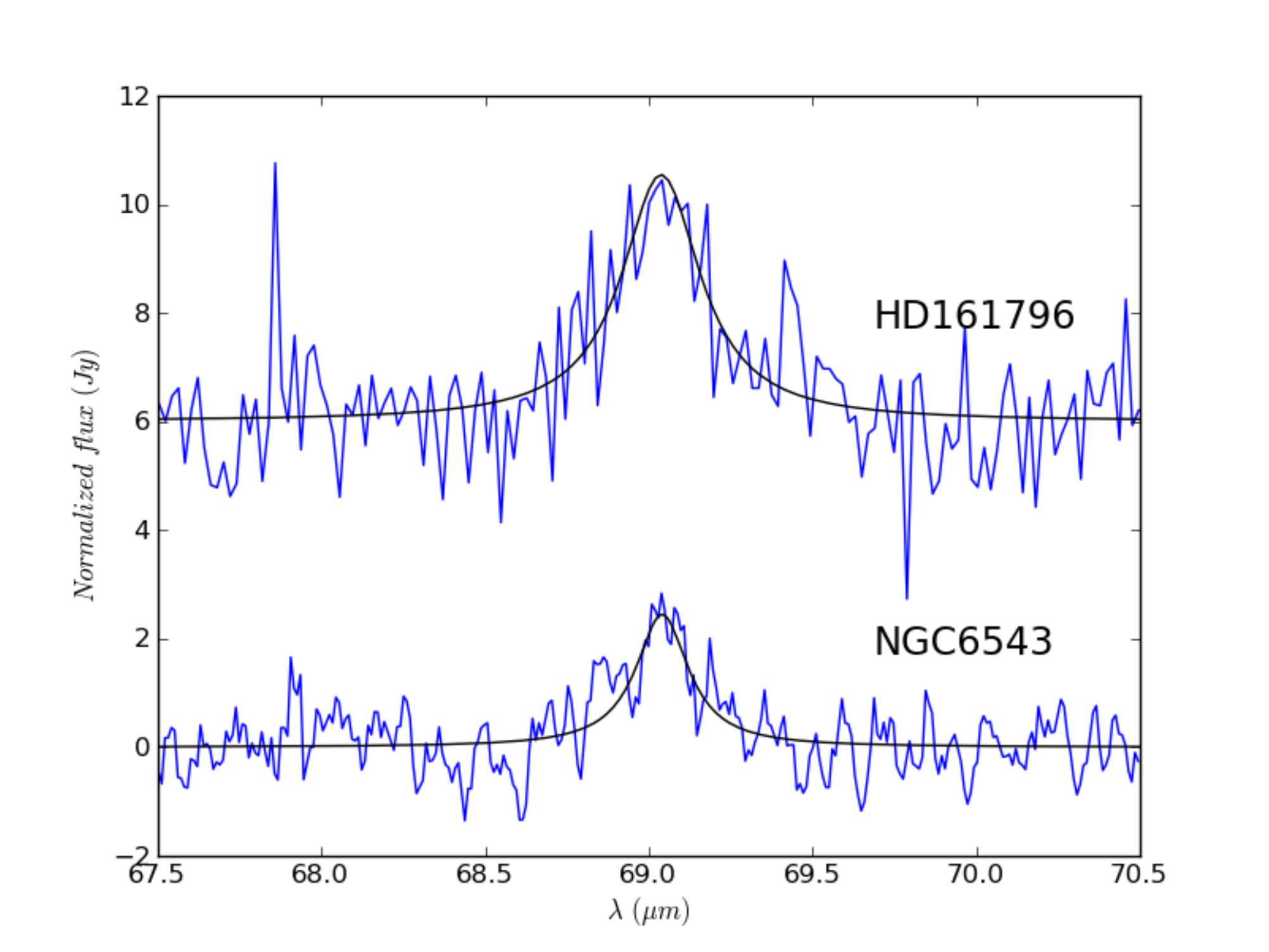}}
 \caption{Fits to the 69\,$\mu$m band of forsterite using temperature dependent opacities
\citep{deVries2011ASPC..445..621D}. Fits are shown for the post-AGB star HD\,161796 and the
planetary nebula NGC~6543. The forsterite dust temperature is lower in HD\,161796  than in NGC 6543.}
\label{Fig:Ben_evolved}
\end{figure}

A solid-state band clearly detected with PACS is the 69\,$\mu$m feature of crystalline olivine (Mg,Fe)$_2$SiO$_4$. The peak and shape of the feature can be used as a dust thermometer, but also to assess the exact composition of the olivine grains (see Fig.~\ref{Fig:Ben_didactiek} and Fig.~\ref{Fig:Ben_evolved}). Combining the 69\,$\mu$m band with the emission peaks at 11.3 and 33.6\,$\mu$m both effects can be disentangled \citep{deVries2011ASPC..445..621D}.

\begin{figure}[htp]
\centering
 \includegraphics[angle=90,width=0.75\textwidth]{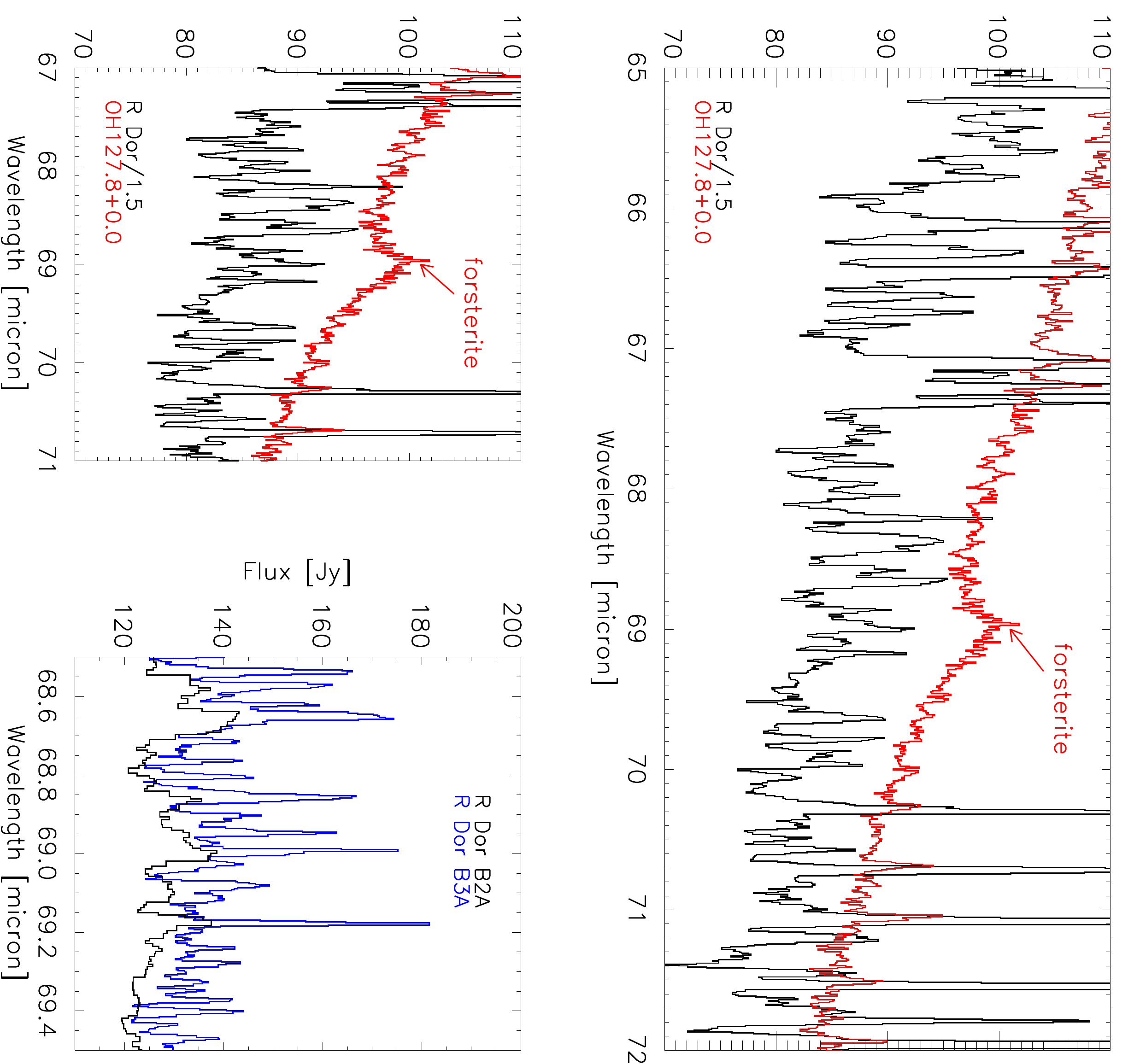}
\caption{Comparison between the 69\,$\mu$m forsterite feature in the oxygen-rich AGB star OH127.8+0.0 \citep[\Mdot\,=\,$5\times10^{-5}$\,\Msun/yr,][]{Lombaert2012} and the rich molecular line spectrum of the oxygen-rich AGB star R~Dor (\Mdot\,=\,$1.2\times10^{-7}$\,\Msun/yr). The upper panel shows the spectra of both targets between 65 and 72\,$\mu$m at a spectral resolution of $\sim$1750. The lower left spectrum zooms into the 69\,$\mu$m region. The lower right figure shows the spectrum of R~Dor around 69 micron in B2A (2nd order, black) at a resolution of $\sim$1750 and in band 3A (3rd order, blue) at a resolution of $\sim$4500. The strongest line peaks in the band 3A spectrum between 68.5-69.25\,$\mu$m are due to ortho- and para-H$_2$O, $^{12}$CO, $^{28}$SiO, and SO$_2$.   }
\label{FIG:OH127_RDor}
\end{figure}

The great advances  hoped to be achieved into this field thanks to the higher sensitivity (and spectral resolution) of \textit{Herschel} PACS/SPIRE compared to ISO LWS are not yet carried through. There are mainly three reasons for this. (1) A firm detection of a (new) solid-state species and a comprehensive analysis of its strength are closely linked to a highly accurate relative flux calibration. Early 2012, the relative calibration accuracy of the PACS spectrometer is typically between 5--10\% (depending on the pointing and possible drifts), which might improve to 1\% when the signal of different spatial pixels can be combined. (2) Thanks to the higher spectral resolution of \textit{Herschel}, it is clear that (molecular) gas lines are contributing to the flux emission at the same wavelengths (see Fig.~\ref{FIG:OH127_RDor}). A dedicated analysis of these gas features is needed to properly determine the strength of the dust emission bands. (3) Most of the solid-state features have the largest extinction 
coefficients in the near to mid-IR wavelength range. 

\subsubsection{Gas-grain reactions} \label{SECT:gas-grain}

The abundance fractions of quite some molecules show a depletion pattern in the dust formation region due to their role in the formation of solid-state species. It is thought that some molecules, as CO, CS and HCN, are quite stable and travel the entire envelope unaltered until they reach the photo-dissocation region of the outer wind, because these molecules do not participate in the formation of dust grains like silicates and corundum \citep{Duari1999A&A...341L..47D}.

\begin{figure}[htp]
\begin{center}
\includegraphics[height=.75\textwidth,angle=90]{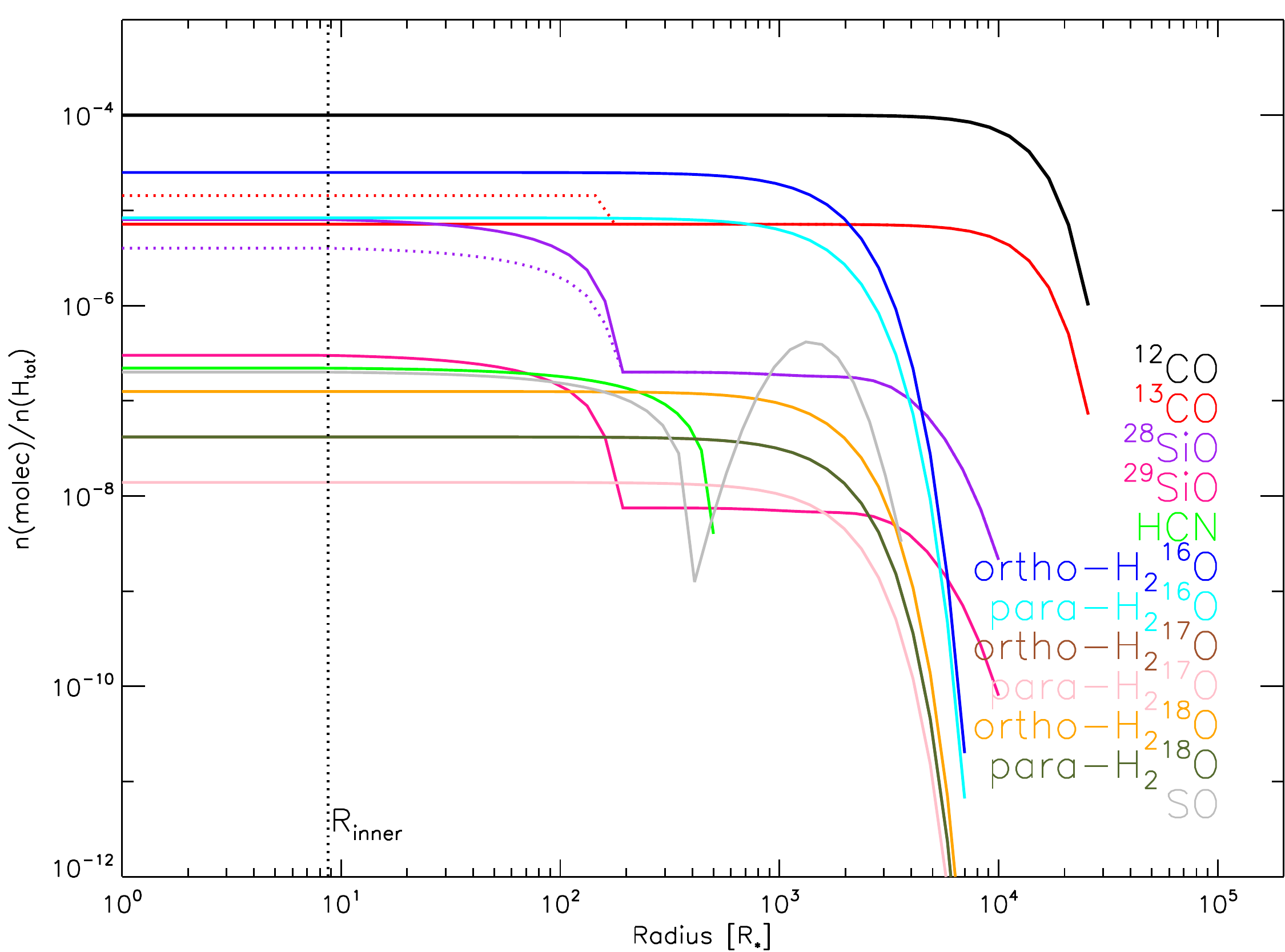}
\vspace*{2ex}
\caption{Fractional abundance stratifications for $^{12}$CO, $^{13}$CO, $^{28}$SiO,  $^{29}$SiO, HCN, SO, ortho-H$_2^{16}$O, para-H$_2^{16}$O, ortho-H$_2^{17}$O, para-H$_2^{17}$O, ortho-H$_2^{18}$O, and para-H$_2^{18}$O as derived by \citet{Decin2010A&A...521L...4D}. For all molecules (except water), the full line represents the results  obtained by \citet{Decin2010A&A...516A..69D}. For  $^{13}$CO and $^{28}$SiO, the new results based on the HIFI data are shown as dotted lines. The fractional abundances for all water isotopologs and isomers are based on HIFI data alone.}
\label{FIG:IKTAu_abundances}
\end{center}
\end{figure}

Thanks to the long wavelength range covered by the three \textit{Herschel} spectrometers different excitation levels of several molecules can be sampled \citep[see, e.g.][]{Decin2010A&A...518L.143D} with the potential to derive possible (molecular) depletion in the envelope. An example is shown for the O-rich AGB star IK~Tau in Fig.~\ref{FIG:IKTAu_abundances}. SiO is clearly depleted in the intermediate wind region, possibly due to the formation of SiO$_2$ (via reaction with OH) whose condensation product is silica and which is also thought to play an active role in the formation of different types of silicates. 

\textit{Herschel} has already observed tens of AGB and supergiant stars. However, only for 3\ of them (IRC\,+10216, $\chi$ Cyg and IK~Tau) a detailed study of the molecular depletion pattern is already performed. This kind of analyses are computationally very expensive and ask quite some dedicated effort of the modeler to analyze tens of lines of different molecules consistently. But even then, we have to realize that this type of results are quite often obtained assuming a spherically symmetric geometry. The \textit{Herschel} images (see Sect.~\ref{SEC:GEOMETRY}), complemented with other optical and sub-millimeter data, show that this is often a too simplistic approach.

\begin{figure}[htp]
\centering
 \includegraphics[width=0.75\textwidth]{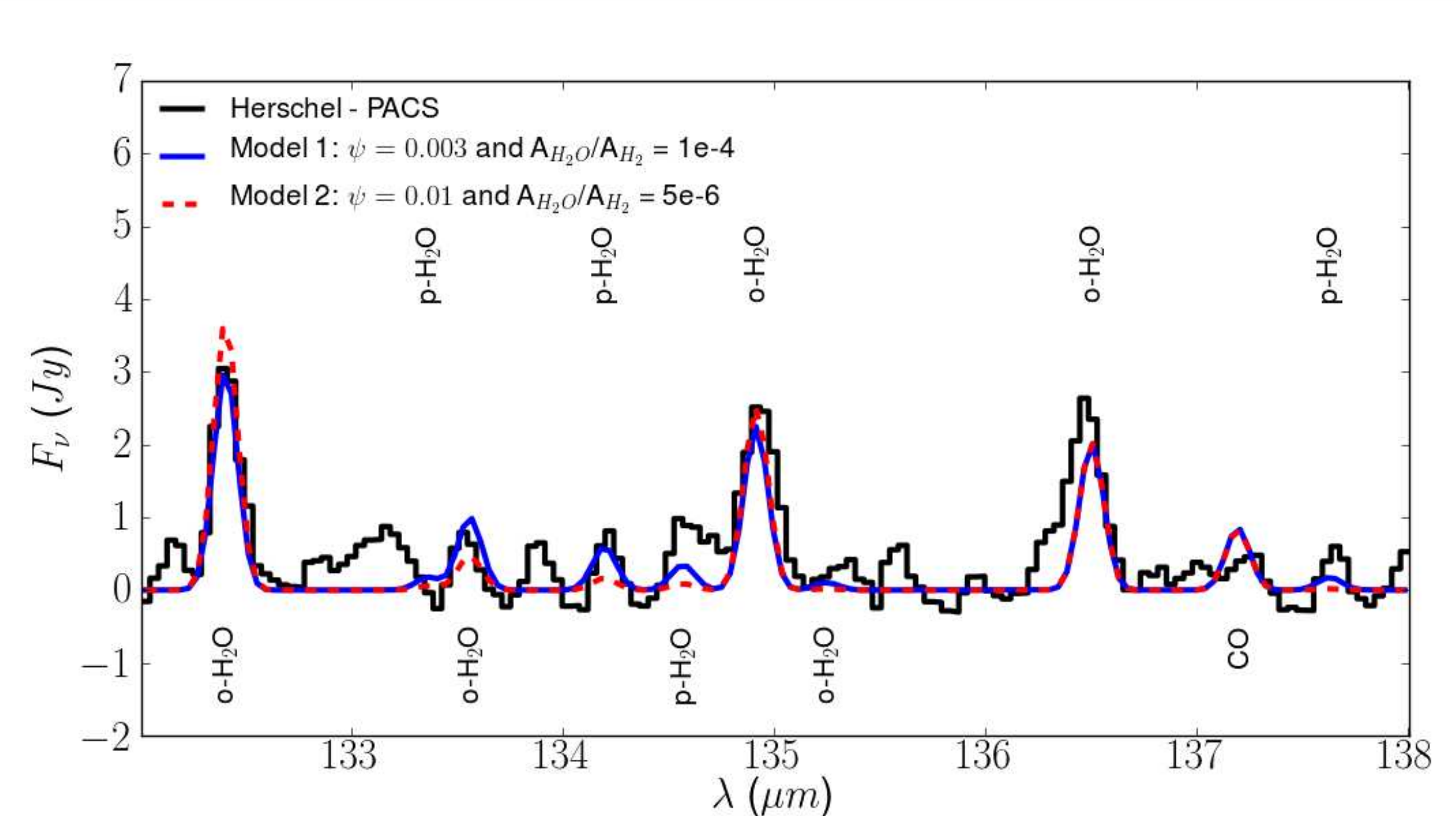}
\caption{\textit{Herschel}/PACS observations of OH127.8+0.0 (black) compared to model predictions. Both a model with a low dust-to-gas ratio $\psi$ of 0.003 and a high water abundance of $1\times10^{-4}$ (blue full line), as well as a model with a high dust-to-gas ratio of 0.01 and a lower water abundance of $5\times10^{-6}$ (red dashed line) can predict the PACS H$_2$O lines equally well.}
\label{FIG:Robin_H2O}
\end{figure}

Another illustrative example of the close interaction between dust and gas is described by \citet{Lombaert2012}. In this paper, it is demonstrated how water vapor lines should be modeled with care, due to a degeneracy between the water abundance and dust-to-gas ratio (see Fig.~\ref{FIG:Robin_H2O}). Water vapor is partly collisionally excited, but for many transitions radiative excitation by the near-IR stellar continuum photons or the thermal dust radiation field are important in establishing the level populations. Extra constraints on the water vapor abundance in the inner envelope can be obtained from the water ice feature at 3.1\,$\mu$m, since the H$_2$O ice fraction leads to a minimum required H$_2$O vapor abundance \citep{Lombaert2012}.

\subsection{Photodissociation of molecules and ion-molecule reactions} \label{SECT:photodissociation}

The penetration of highly energetic interstellar UV and cosmic ray radiation in the tenuous outer envelope induces the photodissociation of  so-called `parent' molecules\footnote{i.e.\ formed in the outer stellar atmosphere or inner envelope and ejected into the intermediate envelope}, which in its turn leads to the formation of new 'daughter' species. E.g., in the outer envelope of both C- and O-rich sources, CN is thought to be produced by the photodissociation of HCN and neutral-neutral reactions \citep{Willacy1997A&A...324..237W, Decin2010A&A...516A..69D}.

The new data obtained with \textit{Herschel} show indications that this process of photodissociation and the production of new molecules due to ion-molecule reactions might even be an important chemical player in the inner envelope. One key discovery made by \textit{Herschel} is the detection of warm water vapor in the inner envelope of the carbon-rich AGB star IRC\,+10216 \citep{Decin2010Natur.467...64D, Neufeld2011ApJ...727L..28N}. The existence of water vapor around carbon-rich AGB stars is remarkable, because, unlike those of oxygen-rich stars, the photospheres of carbon-rich stars are expected to contain very little H$_2$O as they are dominated by CO, HCN, and C$_2$H$_2$ under conditions of thermochemical equilibrium. The water abundances derived from \textit{Herschel} observations of carbon-rich AGB stars are typically 3 to 4 orders of magnitude larger than the expected photospheric abundance. At least in the case of IRC+10216, \textit{Herschel}'s discovery of water in the warm inner envelope imposes 
significant observational constraints on the
various theories that have been proposed. Thus far, at least two possible origins appear to be consistent with the available data. In one picture \citep{Decin2010Natur.467...64D, Agundez2010ApJ...724L.133A}, the water is released by the photodissociation of $^{13}$CO and SiO, thanks to the penetration of interstellar UV radiation through the \textit{clumpy} outflow surrounding the star. In another scenario \citep[][see Sect.~\ref{SECT:non-TE}]{Cherchneff2011A&A...526L..11C}, water results from non-equilibrium chemistry associated with pulsationally-driven shock waves. Other origins including (1) the vaporisation of icy objects (comets or dwarf planets) in orbit around the star \citep{Melnick2001Natur.412..160M}; (2) Fischer-Tropsch catalysis on the surface of small grains \citep{Willacy2004ApJ...600L..87W}; or (3) photochemistry within an outer, photodissociated shell \citep{Agundez2006ApJ...650..374A} could be ruled out since they predict an absence of abundant water within $\sim$100\,AU of the star, which 
is not compliant with the relative strength of the observed higher-excitation
water vapor transitions.

A second important detection has been the widespread presence of water vapor in carbon-
rich AGB stars. A small \textit{Herschel}/HIFI survey for water vapor in eight carbon-rich AGB stars obtained a water detection in every star surveyed, suggesting that IRC\,+10216 is unusual only in its proximity to Earth, and that the presence of water in carbon-rich AGB stars is nearly universal \citep{Neufeld2011ApJ...727L..29N}. In 6 (out of 7) sources already observed with PACS, several low, mid, and sometimes higher excitation H$_2$O lines were detected (see, e.g., Fig.~\ref{FIG:C_stars_H2O}). However, much to our surprise, no water lines were detected in 1 target. While the results of \textit{Herschel} obtained in its first year of observations surprised us in terms of the detection of warm water vapor in carbon-rich stars, the observations taken during the next year implied an inversion of the science question, i.e. `Why don't we see water in all carbon-rich sources?'. The key parameter determining the water abundance is currently not known.

\subsection{CSM-ISM interaction region}

The detection by \textit{Herschel} of the ubiquitous nature of a (turbulent) astropause at the outer edge of the circumstellar envelope stimulates the wish to unravel the chemical composition in the CSM-ISM interaction region. In the end, the thermodynamical and chemical structure in this zone determines the chemical yields injected into the ISM. Currently, the knowledge on the chemical content is very limited. The bow shock regions are either detected in the infrared or far-UV \citep[e.g.][]{Martin2007Natur.448..780M, Ueta2006ApJ...648L..39U, Cox2012A&A...537A..35C}.

The origin of the FUV/NUV emission of the bow shock
around $o$~Cet has been attributed to H$_2$ molecules in the cold gas
which are collisionally excited by hot electron from the postshock
gas \citep{Martin2007Natur.448..780M}. The origin of the FUV/NUV
emission in the interaction zone between the ISM and the stellar
wind of CW Leo has not been modeled in detail, although
it was suggested by \citet{Sahai2010ApJ...711L..53S} that also here
collisionally excited H$_2$ emission might be the cause. The
FUV/NUV emission around CW~Leo peaks at slightly larger distances
from the central target than the PACS/SPIRE infrared flux
excess \citep{Ladjal2010A&A...518L.141L}. Note that not all sources showing a clear IR bow shock detection exhibit UV emission. An illustrative example in this case is $\alpha$~Ori \citep{Decin2012A}.

The infrared emission is  probably due  to thermal emission of cold dust components in the interaction region, with temperatures around 30--150\,K. An attempt to detect probable contribution from low-excitation atomic lines, such as [OI] 63\,$\mu$m and [CII] 158\,$\mu$m, at the IR wavelength bands, has been unsuccessful in case of $\alpha$~Ori \citep{Decin2012B}. Neither is the bow shock detected in low-excitation CO rotational lines. 
Gas temperatures calculated by hydrodynamical models (see Sect.~\ref{SECT:bow shock}) are in the order of few thousand K in the shocked regions. Currently, it is absolutely unclear how this high gas temperature will impact the atomic and molecular abundance pattern in this region. In addition, we still do not know how the turbulence in the astropause will impact the grain temperature.

\section{Some old and new questions} \label{SECT:QUESTIONS}

Since the first detection of a thermally excited circumstellar line, CO $v=0, J=1-0$ at 2.6\,mm, toward IRC\,10216 by \citet{Solomon1971ApJ...163L..53S} our knowledge on the cool envelopes surrounding late-type LIMS has increased steadily. The detection in 1971 was interpreted as being a measure of the radial velocity of a steady mass-loss process, or alternatively, as being the result of an explosive event  which ejected a shell from the central star. Later on, more and more observations corroborated the interpretation of an important mass-loss process during the AGB and supergiant phase. The idea of a steady mass-loss process has remained for a long time, and often  a constant mass-loss rate was assumed in the modeling for the sake of simplicity. That way, the circumstellar envelopes could be modeled using 1D codes. However, astonishing evidence is again given by the \textit{Herschel} observations that these envelopes are far from smooth. Complexity is the rule, not the exception. Since these complex 
structures are the result of a mass-loss process initiated in the upper stellar atmospheres, the upper atmosphere structure comes into question. For supergiants, the huge convective cells detected in the upper atmosphere \citep{Lim1998Natur.392..575L, Chiavassa2010A&A...515A..12C} are indeed the direct proofs of non-homogeneous atmospheric structures. Cool gas should co-exist with the hotter gas. Related to this convective motion, is the idea of a solar-like magnetic activity cycle \citep{Soker2000ApJ...540..436S}. Owing to the appearance of magnetic cool spots during the active phase, the gas temperature can be locally reduced. This can enhance dust formation and thus lead to a higher mass-loss rate from magnetic cool spots. For the less massive counterparts, the AGB stars, radiation pressure on dust grains is still thought to be the main driver of the mass-loss process. However, in that case, one would expect a more or less smooth envelope structure. The complex inner envelope structures shown in Fig.~\ref{FIG:inner_envelope} prove that another physical process, of which the strength varies across the upper stellar atmosphere and/or inner envelope, can not be neglected. A possible suggestion is related to dust shadowing or multi-scattering processes, but the data currently lack spatial resolution to corroborate this idea. How this affects the  wind acceleration should still be sorted out. 

We also should realize that stellar structure and evolution models are still using a simplified (analytical) description of the mass-loss rate. The \textit{Herschel} observations show that one might start wondering about  this type of assumptions. Improving this description is important since the mass-loss rate determines the fraction of stars which finally will end their live as supernovae or white dwarfs.

In my opinion, one of the main answers we can get from the \textit{Herschel} observations concerns the \textit{empirical} determination of the global envelope structure, i.e.\ temperature $T(r)$ and abundance structure $\epsilon(r)$. Thanks to the extended wavelength coverage of the three instruments, we are sampling the CO ladder from $J=4-3$ to $J=53-52$, even in some rare cases seeing rotational transitions in vibrationally excited states. Complemented with some ground-based data (to get the lower rotational transitions), these CO data form an excellent thermometer. Last few years, effort has been spent to derive the temperature structure from first principles \citep[see, e.g.,][]{Decin2006A&A...456..549D}. However, still quite some assumptions are part of the functions determining the cooling and heating rates per unit volume. A  more reliable temperature structure can be derived from the data themselves. The same holds for the abundance structure. The \textit{Herschel} spectra are extremely rich, 
especially in terms of molecular line transitions. In the case that different excitation regimes are sampled $\epsilon(r)$ can be derived, yielding observational constraints for chemical models. For the first time, infrared and submillimeter data of a large set of evolved LIMS, covering different (circum)stellar parameters, is available to the community. A homogeneous analysis of these data will form an important testbed for the chemical models. The observation of water vapor in carbon-rich stars can serve as a good example. While in 2001, it was argued that water vapor in this type of carbon-rich environment was due to the vaporization of icy bodies \citep{Melnick2001Natur.412..160M}, the detection of high-excitation water vapor lines by \textit{Herschel} gave arguments against this interpretation \citep{Decin2010Natur.467...64D}. At that moment, the idea of photo-induced chemistry in a non-homogeneous environment was postulated. Later on, updated chemistry models showed that pulsationally-induced non-
equilibrium 
chemistry could explain the observations as well \citep{Cherchneff2011A&A...526L..11C}. And while in 2010, we could prove the ubiquitous nature of water vapor in C-rich envelopes, in 2012 we start wondering why we do not see water vapour in \textit{all} C-rich AGB stars (see Sect.~\ref{SECT:photodissociation}). This proves that in the last $\sim$10 years time our ideas on the chemistry in these envelopes have changed drastically.

A new field nicely uncovered by \textit{Herschel} concerns the CSM-ISM interaction region. The data shown by \citet{Cox2012A&A...537A..35C} show for the first time a whole zoo of interaction phases, roughly sub-divided in four categories. The complexity at the contact discontinuity is huge, the dynamical and temperature structure still largely unknown, let alone for the chemical pattern. How does this harsh environment influence the structure of the grains and the photodissociation of the molecules? And how does this energetic environment determine the composition of the chemical yields enriching the ISM?

\section{Future perspectives} \label{SECT:PERSPECTIVES}

\begin{figure}[htp]
\centering
 \includegraphics[angle=180,width=0.75\textwidth]{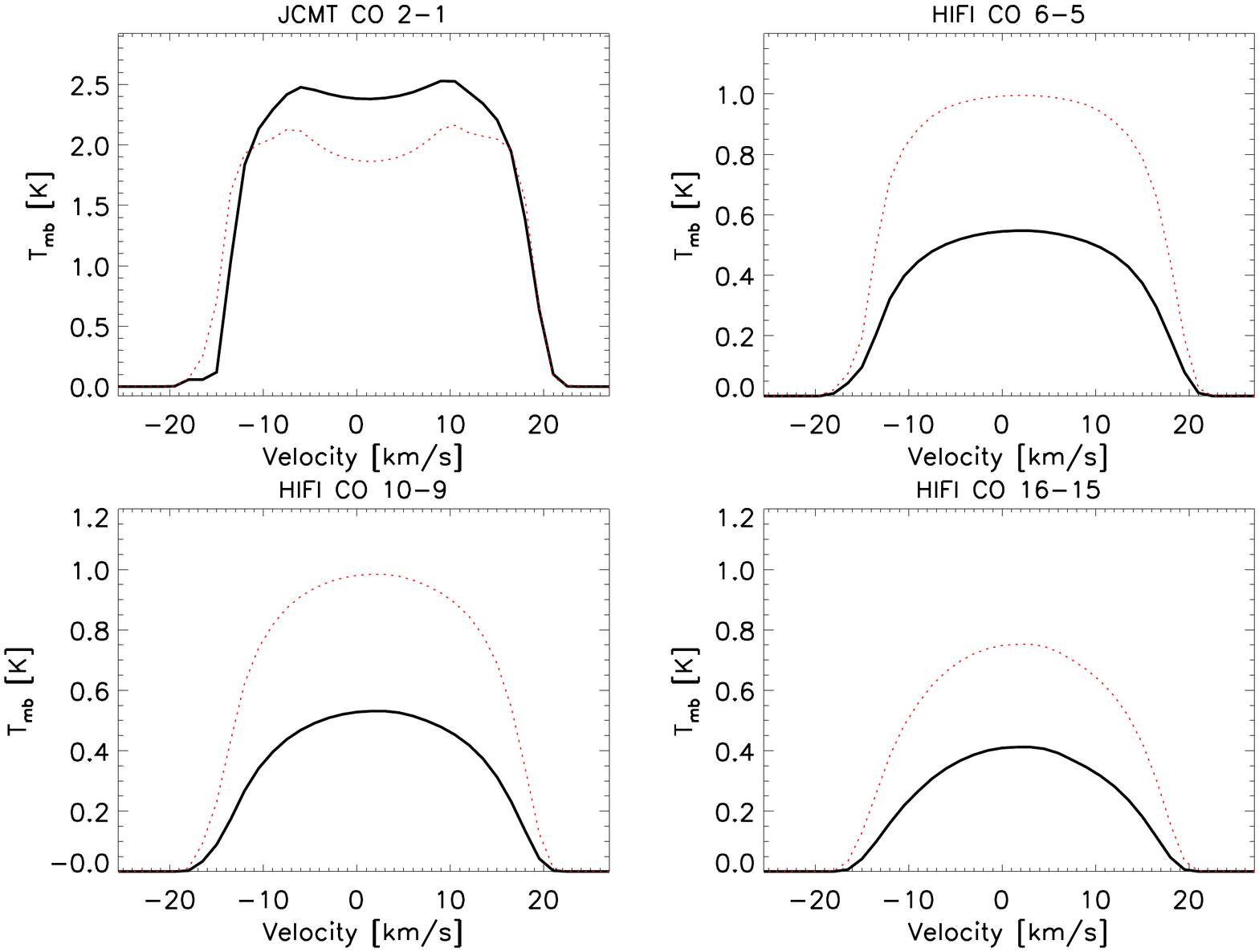}
\caption{Comparison between theoretical spectra for different CO transitions for a model with parameters similar to the O-rich giant IK~Tau \citep{Decin2010A&A...516A..69D}. The full black line represents a model for  a luminosity of 1000\,\Lsun (\Teff=2200\,K, \Rstar=$1.5\times10^{13}$\,cm), the red dotted line is the result for a model with a luminosity of 2000\,\Lsun, simulated by increasing the effective temperature, \Teff, to 2600\,K, and lowering the stellar radius to  \Rstar=$1.3\times10^{13}$\,cm. A change in luminosity mainly affects the higher-excitation lines formed closer to the central star.}
\label{FIG:compare_L}
\end{figure}

In my opinion, two keywords on the exploitation of the \textit{Herschel} in the next 10 years are 'confrontation' and 'consistency', if possible coupled to each other. We are still in the first exploration status of the \textit{Herschel} data, scrolling through images and spectra and looking for spectacular results. These first results are important in their own right, and do give us already some nice clues on  late stages of stellar evolution. A next step should aim for an in-depth confrontation of the data with current theoretical models. On the one hand, this will demonstrate the strength of the models, and show how some general quantities might be extrapolated to other evolved stars, maybe in other galaxies. On the other hand, this will inevitably show us the limitations of the models/our knowledge. One of the main limitations the field of evolved-stars research is currently confronted with concerns 'consistency' between applied models. There are models focusing on the gas chemistry or on grain chemistry,
 models solving the thermodynamic equations, models focusing on multi-D radiative transfer etc. However, none of the current models is fully consistent. The dust grains and gas molecules are intricately interacting in the CSEs, both radiatively and dynamically. Therefore it is of interest to adopt a model that is as self-consistent as possible. One aspect concerns the fact that AGB stars are dynamically unstable, resulting in the star changing its luminosity when going from maximum to minimum light. This severely affects its spectral energy distribution, as shown by e.g.\ \citet{DeBeck2012}. However, not only the thermal dust emission is affected, but a change in luminosity (and hence dust opacity) will change the line strengths of quite some lines, both in the infrared \citep[see, e.g.,][]{vanMalderen2003PhDT..........V} and sub-millimeter regime \citep[see, e.g.,][and Fig.~\ref{FIG:compare_L}]{Lombaert2012}. This effect should be taken into account when modeling data taken at different phases.

Another example concerns the formation of some gas species (as H$_2$O), which might influence the thermal structure (cooling or heating the envelope). This local temperature change might affect in its turn the formation of specific dust grains, which are known to act as wind drivers. Added to this are effects as backwarming or dust shadowing in combination with, e.g., multi-scattering of stellar and dust-emitted photons. This might yield non-isotropic dust formation. Admittedly, this type of `consistent' models are CPU consuming, but the theoretical knowledge on each specific sub-field is there. It just should be combined. While  \textit{Herschel} gives us a whole zoo of infrared and submillimeter data, the theoretical models are currently lagging behind.

Another issue, with which the \textit{Herschel} data are confronting us, is our limited knowledge on the chemical properties of many gas and dust species. Even in case of relatively `simple' molecules, as C$_2$H our radiative transfer treatment is still very limited: C$_2$H is a linear molecule with an electronic $^2\Sigma^+$ ground-state configuration of which nowadays radiative ransfer calculations are done assuming a $^1\Sigma$ approximation. The dipole moments of the fundamental vibrational states of C$_2$H are still unknown \citep{DeBeck2012}. This lack of knowledge might severely limit the accuracy of the derived results. This has been demonstrated in the case of  H$_2$O, of which it is now well established that one overestimates the H$_2$O abundance by an order of magnitude in the case one neglects the infrared pumping of H$_2$O via various vibrational bands \citep{Agundez2006ApJ...650..374A}. This situation will hopefully improve in the near future for simple 2 and 3-atomic molecules thanks to 
initiatives as the Cologne Database for Molecular Spectroscopy \citep[CDMS,][]{Muller2005JMoSt.742..215M}\footnote{http://www.astro.uni-koeln.de/cdms/}, the JPL catalog \citep{Pickett1998JQSRT..60..883P}\footnote{http://spec.jpl.nasa.gov}, {\sc basecol} \citep{Dubernet2006JPlR....7..356D}\footnote{http://basecol.obspm.fr/}, etc.

\section{Conclusions} \label{SECT:Conclusions}

To conclude, the harvest by the \textit{Herschel} observations in the field of evolved stars is large. New ideas on the chemical, thermal and dynamical structure in the envelopes created by the dominant mass-loss process have emerged. As of today, this is reflected in more than 150 refereed-journal papers in the field of evolved AGB and supergiants stars containing \textit{Herschel} data. Quite some evolved stars are scheduled to be observed with \textit{Herschel}. The comprehensive analyses and consistent confrontations with theoretical models will still take years.

A next \textit{observational} step, which has already begun, concerns high spatial resolution observations, as will be provided by instruments as ALMA. With its high sensitivity, ALMA will enable us to resolve the inner wind structures of the most nearby targets and the high-resolution maps of the winds will be an excellent tracer of the dust and gas distributions, and will in that way complement the extremely valuable \textit{Herschel} data.


\section*{Acknowledgements}
LD wants to thank all the colleagues in the \textit{Herschel} PACS, SPIRE and HIFI consortia for the fruitful discussions and nice collaborators. Special thanks to Robin Lombaert, Ben de Vries, Allard Jan Van Marle and Nick Cox for providing me with illustrative figures.
 This research has made use of NASA's Astrophysics Data System Bibliographic Service and the SIMBAD database, operated at CDS, Strasbourg, France.
PACS has been developed by a consortium of institutes led by MPE (Germany) and including UVIE (Austria); KUL, CSL, IMEC (Belgium); CEA, OAMP (France); MPIA (Germany); IFSI, OAP/AOT, OAA/CAISMI, LENS, SISSA (Italy); IAC (Spain). This development has been supported by the funding agencies BMVIT (Austria), ESA-PRODEX (Belgium), CEA/CNES (France), DLR (Germany), ASI (Italy), and CICT/MCT (Spain). SPIRE has been developed by a consortium of institutes led by
Cardiff Univ. (UK) and including Univ. Lethbridge (Canada);
NAOC (China); CEA, LAM (France); IFSI, Univ. Padua (Italy);
IAC (Spain); Stockholm Observatory (Sweden); Imperial College
London, RAL, UCL-MSSL, UKATC, Univ. Sussex (UK); Caltech, JPL,
NHSC, Univ. Colorado (USA). This development has been supported
by national funding agencies: CSA (Canada); NAOC (China); CEA,
CNES, CNRS (France); ASI (Italy); MCINN (Spain); SNSB (Sweden);
STFC (UK); and NASA (USA). HIFI has been designed and built by a consortium of institutes and
university departments from across Europe, Canada and the United States
under the leadership of SRON Netherlands Institute for Space Research,
Groningen, The Netherlands and with major contributions from Germany,
France and the US.  Consortium members are: Canada: CSA, U.Waterloo;
France: CESR, LAB, LERMA, IRAM; Germany: KOSMA, MPIfR, MPS; Ireland,
NUI Maynooth; Italy: ASI, IFSI-INAF, Osservatorio Astrofisico di
Arcetri- INAF; Netherlands: SRON, TUD; Poland: CAMK, CBK; Spain:
Observatorio Astron\'omico Nacional (IGN), Centro de Astrobiolog\'{\i}a
(CSIC-INTA); Sweden: Chalmers University of Technology - MC2, RSS \&
GARD; Onsala Space Observatory; Swedish National Space Board, Stockholm
University - Stockholm Observatory; Switzerland: ETH Zurich, FHNW; USA:
Caltech, JPL, NHSC. HCSS / HSpot / HIPE is a joint development by the
Herschel Science Ground Segment Consortium, consisting of ESA, the NASA
Herschel Science Center, and the HIFI, PACS and SPIRE consortia.

\vspace*{-1ex}
\vspace*{-.5cm}
\fullreferences
\end{document}